\documentclass[preprint,showpacs,preprintnumbers,amsmath,amssymb]{revtex4}

\usepackage{graphicx}
\usepackage{dcolumn}
\usepackage{bm}
\usepackage{amsmath}

\begin{document}

\title{Evidence for mass zeros of the fermionic determinant in four-dimensional quantum electrodynamics }

\author{M. P. Fry\footnote{mpfry@maths.tcd.ie}}

\affiliation{School of Mathematics, University of Dublin, Trinity College, Dublin 2, Ireland}

\date{\today}

\begin{abstract}
The Euclidean fermionic determinant in four-dimensional
quantum electrodynamics is considered as a function of the
fermionic mass for a class of $O(2)\times O(3)$ symmetric background
gauge fields. These fields result in a determinant free of
all cutoffs. Consider the one-loop effective action, the
logarithm of the determinant, and subtract off the
renormalization dependent second-order term. 
Suppose the small-mass behavior of this remainder is fully determined by the chiral anomaly.  
Then either
the remainder vanishes at least once as the fermionic mass
is varied in the interval $0 < m < \infty$  or it reduces to its
fourth-order value in which case the new remainder, obtained
after subtracting the fourth-order term,  vanishes at least
once. Which possibility is chosen depends on the sign of simple integrals involving the field strength tensor and its dual.  
\end{abstract}

\maketitle

\section{Introduction}
Within the standard model fermionic determinants are required
for the calculation of every physical process. These determinants
produce an effective functional measure for the gauge
fields when the fermionic fields are integrated. They are the means
by which virtual fermion loops are incorporated into a calculation.
Without them color screening, quark fragmentation into hadrons and
unitarity would be lost. In quantum electrodynamics the biggest barrier to understanding
its nonperturbative structure is its fermionic determinant. These
determinants are therefore fundamental.

They are also hard to calculate and physicists by and large lost
interest in them during the 1980s. With the advent of large machines
lattice QCD physicists are now starting to include the determinant
in their calculations. Analytic results for QCD and QED determinants
are very scarce, especially in four dimensions. Such results as they
become available will serve as benchmarks for determinant algorithms,
including the various lattice
 discretizations of the Dirac operator in use, and
hence a means of reliably estimating computational error, a major
problem in lattice QCD at present.
    
Most analytic nonperturbative results obtained so far deal with
the dependence of the determinant on the coupling constant. Little
attention has been given to their dependence on the fermion's mass.
One notable exception is the work of Dunne et al. \cite{bib_1}, 
which
gives a semi-analytic calculation of the QCD determinant's mass
dependence in an instanton background.

In two-dimensional Euclidean QED the author has shown that
mass can have a profound effect on its determinant. Namely, for a
large class of centrally symmetric, finite-range background gauge
fields the growth of the determinant in the limit $mR << 1$  followed by
$|e \Phi|  >>1$ is

\begin{equation}
\ln\det \sim - \frac{|e \Phi|}{4 \pi} \ln \left ( \frac{|e\Phi|}{(mR)^2} \right ),
\label{equation 1}
\end{equation}
where $\det$ denotes the determinant, $R$ is the field strength's range, and
$\Phi$ is the background field's flux \cite{bib_2}. 
In the massless case, the
Schwinger model, the determinant is quadratic in the field strength.

The second example of the nontrivial mass dependence of $\det$ in
Euclidean QED$_2$  is the presence of mass zeros. Let $\det$ be written as

\begin{equation}
\ln\det =  \Pi_2   + {\ln\det}_3,
\label{equation 2}
\end{equation}
where $\Pi_2$ is the second-order vacuum polarization graph and $\ln\det_3$ is a technical term, defined
in Sec. II, for the remainder after the conditionally convergent second-order
term has been isolated and made gauge invariant by some regularization
procedure. Then there is at least one real value of $m$  at which
$\ln\det_3  = 0$ when $0<  |e \Phi |  <2\pi $ , subject to some mild restrictions on the field strength \cite{bib_3}.
There may be other mass zeros. Now
recall Schwinger's result \cite{bib_4} 
that $\ln\det_3  = 0$ when $m  = 0$. For
fields with   $\Phi = 0$ then it is also true that \cite{bib_5}

\begin{equation}
                      \lim_{m = 0} {\ln\det}_3  = 0;
\label{equation 3}
\end{equation}
otherwise not. So the result is this: when $0< |e\Phi|   < 2\pi$   the zero in
$m$ of $\ln\det_3$  moves up from $m  = 0$ to some finite value $m > 0$.
Beyond $|e \Phi|  > 2\pi$  we can say nothing definite yet.

%
%

The obvious question to ask is whether there are mass
zero(s) in the remainder term of $\ln \det$ in QED$_4$  , denoted by
$\ln\det_5$   . The background gauge fields $A_\mu (x)$ considered in two
dimensions have a slow $1/|x|$ falloff resulting in a nonvanishing chiral anomaly $\Phi / 2\pi$. Here we will consider
a large class of $O(2) \times O(3)$ symmetric background gauge fields that
also have a $1/|x|$ falloff with a nonvanishing chiral anomaly.
If the small-mass behavior of the remainder is fully determined
by the chiral anomaly, as in two dimensions, then there are circumstances in which
mass zeros are present in the remainder. The idea of the proof
is extremely simple: show that for $m \rightarrow 0$ the remainder is
negative and that as $m \rightarrow \infty$  it becomes positive. The demonstration
that the chiral anomaly determines the small-mass behavior of
the remainder turns out to be nontrivial, and we are not able to
settle this matter here. Evidence is presented that it does, but
it is not conclusive.

At this point it may be asked why these mass zeros
    for a special class of background gauge fields are of
    interest. First and foremost they are a truly nonperturbative
    result for the exact QED$_4$ determinant. As such, they
    would serve as a benchmark result that lattice theorists
    could aim to reproduce. As discussed in Sec. II, once the
    second- and fourth-order contributions to $\det_{{ren}}$ are
    isolated the remainder of $\det_{{ren}}$ is determined by the
    distribution of its complex zeros in the coupling
    constant plane. Little is known about how these zeros
    distribute themselves. The presence of mass zeros in the
    remainder terms in $\ln \det_{{ren}}$ must 
place a strong constraint
    on their distribution which future work could deal with.

    In Sec. II $ \det_{{ren}}$    is defined and some of its properties are
reviewed. Section III introduces the background gauge fields used
in the calculation. Section IV is an introduction to the zero-mass
limit of the remainder and some of the subtleties involved.
Section V establishes that all of the square-integrable zero modes
of the Dirac operator $\displaystyle{\not}D$ have positive chirality. In addition it is
necessary to know the scattering states and low-energy phase
shifts associated with the background gauge field, and this is
done in Sec. VI. Section VII gives an analysis of the low energy
behavior of the exact negative chirality propagator and seeks to
justify a particular approach to proving that the chiral anomaly
is sufficient to describe the small-mass limit of the remainder.
Section VIII demonstrates that the remainder can become positive
as $m \rightarrow \infty$. Finally, Sec. IX summarizes our conclusions.

\section{QED$_4$  Determinant}

We begin by reviewing some established results for the QED$_4$
determinant \cite{bib_6, bib_7}. 
By fermionic determinant we mean the ratio of
determinants of the interacting and free Euclidean Dirac operators,
$ \det (\displaystyle{\not}P - e\displaystyle{\not}A + m) / \det (\displaystyle{\not}P + m)$, defined by the renormalized determinant
on $\mathbb{R}^4$, namely
\begin{equation}
{\det}_{{ren}}   = \exp(\Pi_2  +\Pi_3   +\Pi_4   ){\det}_5 (1 - eS\displaystyle{\not}A),       
\label{section2 equation 1}
\end{equation}
where
\begin{equation}
{\ln\det}_5  = \mbox{Tr} \left [ \ln (1 - eS\displaystyle{\not}A) + \sum_{n=1}^4 \frac{(eS\displaystyle{\not}A)^n} {n} \right ],    
\label{section2 equation 2}
\end{equation}
and $S = (\displaystyle{\not}P + m)^{-1}  ;  \Pi_{2,3,4}$      are the second, third and fourth-order
contributions to the one-loop effective action defined by some
consistent regularization procedure together with a charge
renormalization subtraction in    $\Pi_2$ . The regularization should also
result in    $\Pi_3 = 0$  by C-invariance, and it should give a gauge-invariant result for  $\Pi_4$  . The remainder, $\det_5$, after these
subtractions is gauge invariant and has a well-defined power series
expansion without regularization. The remainder ${\ln\det}_3$  in (\ref{equation 2})
is given by (\ref{section2 equation 2})  with the restriction $n = 1,2$.

   The operator $S\displaystyle{\not}A$ is a bounded operator on
the Hilbert space $L^2 (\mathbb{R}^4  ,  \sqrt{k^2  + m^2}  d^4 k)$ for 
$A_\mu\hspace{-4mm}\in\hspace{-4mm}\bigcap\limits_{n>4}   L^n (\mathbb{R}^4)$, in which
case it belongs to the trace ideal $\mathcal{C}_n$   for $n>4$ $\left[\mathcal{C}_n = \{K|\mbox{Tr}(K^{\dagger} K)^{\frac{n}{2}}  < \infty\} \right]$
\cite{bib_6, bib_7, bib_8, bib_9}. 
This includes the case when $A_\mu (x)$  falls off as ${1}/{|x|}$ as $|x|  \rightarrow \infty$.
As a result $\det_5$  is an entire function of the coupling $e$, and it can be
represented in terms of the discrete complex eigenvalues ${1}/{e_n}$ of the
non-Hermitian compact operator $S\displaystyle{\not}A$ \cite{bib_10}:
\begin{equation}
        {\det}_5  (1 - eS\displaystyle{\not}A) =   \prod_n \left [ \left (1 - \frac{e}{e_n} \right )\exp\left (\sum_{k=1}^4   \frac{({e}/{e_n} )^k}{k}\right ) \right ].  
\label{section2 equation 3}
\end{equation}

By C-invariance and the reality of ${\det}_5$    these eigenvalues appear
in quartets $\pm e_n  , \pm \bar{e}_n$  or as imaginary pairs. Because ${\det}_{{ren}}$   has no
zeros for real $e$ when $m \not= 0$ \cite{bib_11} 
and ${\det}_{{ren}}(e=0) = 1$, it is positive
for real $e$. Because $S \displaystyle{\not}A \in \mathcal{C}_n$,      $n>4$, it is of order $4$. 
This means that for suitable positive constants $A(\epsilon)$, $K(\epsilon)$ and any complex value of $e$,
$| {\det}_{{ren}} | < A(\epsilon)\exp\left ( K(\epsilon )|e|^{4+\epsilon}   \right )$   for any $ \epsilon  >0$.
The first paper to show that  ${\det}_{{ren}}$ is of order 4 was that
in \cite{bib_25}.

In the coordinate space representation of the operator $S(P)\displaystyle{\not}A(X)$,
the propagator is given by
\begin{equation}
                S(x) = \frac{m}{4 \pi^2} \left ( i \displaystyle{\not}\partial + m \right ) \frac{K_1 (m |x|)}{|x|}.
\label{section2 equation 4}
\end{equation}
Here $S$ is an analytic function of $m$ throughout the complex $m$-plane
cut along the negative real axis. A theorem of Gohberg and Kre\u{\i}n
\cite{bib_12}
states that if $A(\mu ) \in \mathcal{C}_1$    and is analytic in $\mu$   in some region
then so is ${\det}(1-A(\mu ))$.
In our case $S\displaystyle{\not}A \in \mathcal{C}_{4+\epsilon}$     , requiring the four
subtractions from the logarithm in (\ref{section2 equation 2}). These subtractions are
easily incorporated into Gohberg and Kre\u{\i}n's proof for $S\displaystyle{\not}A \in \mathcal{C}_1$     ,
provided use is made of the inequality \cite{bib_10, bib_13},
\begin{equation}
|{\det}_n ( 1 + A ) | \leq e^{\Gamma_k ||A ||_n^n},
\label{section2 equation 5}
\end{equation}
for $A \in \mathcal{C}_n$, $||A||_n^n  = \mbox{Tr}|A^\dagger A|^{\frac{n}{2}}$, and  $\Gamma _k$  is a constant. Therefore $\det_5 (1-S\displaystyle{\not}A)$ is
infinitely differentiable in $m$ on the interval $(0,\infty  )$.

    The regularization procedure used here is Schwinger's heat kernel
representation \cite{bib_14}:
\begin{eqnarray}
{\ln\det}_{{ren}} &=& \begin{matrix}\frac{1}{2}\end{matrix} \int_0^\infty \frac{dt}{t} \int d^4 x \left \{ \mbox{tr} \langle x| e^{-P^2t} - e^{-(D^2 + \frac{1}{2} \sigma F)t}|x\rangle
+ \frac{1}{24\pi^2} F^2_{\mu \nu} (x) \right \} e^{-tm^2} \nonumber\\ \nonumber\\
&=&\frac{1}{8\pi^2} \int \frac{d^4 k}{(2\pi)^4} | \hat{F}_{\mu \nu} (k) |^2 \int_0^1 dz z (1 -z ) \ln \left ( \frac{z ( 1 - z) k^2 + m^2 } { m^2} \right )\nonumber\\
\nonumber\\
& & + \Pi_4  + {\ln\det}_5 (1 - S \displaystyle{\not}A).
\label{section2 equation 6}
\end{eqnarray}
Here $e$ has been absorbed into $A_\mu$  , $D^2  = (P-A)^2$ , $ \sigma_{\mu \nu}  = [\gamma_\mu  ,\gamma_\nu  ]/2i$,
$\gamma^\dagger_\mu  = -\gamma_\mu$   , $\hat{F}_{\mu \nu}$   denotes the Fourier transform of $F_{\mu \nu}$  , and $m$ is the
fermionic mass. A second-order on-shell charge renormalization
subtraction has been incorporated. All terms appearing on the
right-hand side of (\ref{section2 equation 6}) follow from the heat kernel expression on
the left-hand side. The requirement that $A_\mu \in \bigcap\limits_{n > 4} L^n ( \mathbb {R}^4)$  and certain
differentiability conditions on $A_\mu$  introduced later are sufficient to 
ensure that (\ref{section2 equation 6}) makes mathematical sense.

    In the representation
\begin{eqnarray}
\gamma_0 = -i \left ( \begin{array}{cc} 0 & 1 \\ 1 & 0 \end{array} \right )\mbox{, }
\gamma_k =  \left ( \begin{array}{cc} 0 & \sigma_k \\ -\sigma_k & 0 \end{array} \right )\mbox{, } 
\gamma_5 = \gamma_0 \gamma_1 \gamma_2 \gamma_3 =  \left ( \begin{array}{cc} 1 & 0 \\ 0  & -1\end{array} \right ),  
\label{section2 equation 7}
\end{eqnarray}
one gets
\begin{equation}
D^2 + \begin{matrix}\frac{1}{2}\end{matrix} \sigma F =\left( \begin{array}{cc} H_+ & 0 \\ 0 & H_- \end{array} \right ) , 
\label{section2 equation 8}
\end{equation}
where
\begin{equation}
H_\pm = (P - A)^2 - {\boldsymbol{\sigma}} \cdot (\mathbf{{B}}\pm \mathbf{{E}}).
\label{section2 equation 9}
\end{equation}

    Denote the remainder in (\ref{section2 equation 6}) after removing the renormalization
dependent second-order term by
\begin{equation}
\mathcal{R} = \Pi_4 + {\ln\det}_5 (1 - S \displaystyle{\not}A). 
\label{section2 equation 10}
\end{equation}
It will be shown in Sec.\hspace{-2mm} V that all the zero modes of the Dirac operator
are confined to the positive chirality sector for the class of
$O(2)\times O(3)$ symmetric background fields used to calculate $\det_{{ren}}$.
Differentiating (\ref{section2 equation 6}) with respect to $m^2$   allows one to isolate $H_+$.
After some rearrangement of terms there follows
\begin{eqnarray}
m^2 \frac{\partial \mathcal{R}}{\partial m^2} &=& \begin{matrix}\frac{1}{2}\end{matrix} m^2 \mbox{Tr} \left[ (H_+ + m^2)^{-1} - (H_- + m^2)^{-1} \right] \nonumber\\
\nonumber \\
& & + m^2 \int_0^\infty dt e^{-t m^2} \int d^4 k \left \{ \mbox{tr} \langle k| e^{-tH_-} - e^{-t P^2} |k\rangle \right. \nonumber\\
\nonumber \\
&& \left. - \frac{1}{128\pi^6} | \hat{F}_{\mu \nu} (k)|^2 \int_0^1 dz z (1 - z) e^{-k^2z (1 -z) t } \right \},
\label{section2 equation 11}
\end{eqnarray}
where the spin traces are now over $2\times2$ matrices. Equation (\ref{section2 equation 11}) expanded
as a power series defines the conditionally convergent fourth-order
term in $\mathcal{R}$. This requires iterating the second term in (\ref{section2 equation 11}) four
times using the operator identity
\begin{eqnarray}
e^{-t(P^2+ V) } - e^{-t P^2} & = & -\int_0^t ds e^{- (t-s)(P^2 + V) } V e^{-sP^2}, 
\label{section2 equation 12}
\end{eqnarray}
with $V = -AP - PA + A^2   + \boldsymbol{\sigma} \cdot  (\mathbf{{E}}-\mathbf{{B}})$. The result is
\begin{eqnarray}
m^2 \frac{\partial \mathcal{R}}{\partial m^2} &=& \begin{matrix}\frac{1}{2}\end{matrix}  m^2 \mbox{Tr} \left[  (H_+ + m^2)^{-1} - (H_- + m^2)^{-1} \right ] \nonumber \\
\nonumber \\
&-& \frac{m^2}{16\pi^2} \int \frac{d^4k}{(2\pi)^4} \frac{ \mathbf{\hat{E}}(k)\cdot \mathbf{\hat{B}} (-k) + \mathbf{\hat{E}}(-k) \cdot \mathbf{\hat{B}} (k) } { z ( 1 - z) k^2 + m^2 } \nonumber \\
\nonumber \\
&+& m^2 \frac{\partial}{\partial m^2} \Pi_4^- \nonumber \\
\nonumber \\
&-& m^2  \mbox{Tr}\Big  [ \Delta_- V \Delta V \Delta V \Delta V \Delta V \Delta \nonumber\\
&&-  ( \Delta A^2 \Delta V \Delta V \Delta V \Delta  + \mbox{ all perms. of $A^2$, $V$ } ) \nonumber \\
&&+  \Delta A^2 \Delta A^2 \Delta A^2 \Delta \Big]. 
\label{section2 equation 13}
\end{eqnarray}
The second term in (\ref{section2 equation 13}) is the remainder after adding the second-order contribution from the second term in (\ref{section2 equation 11}) to the last term.  This remainder would be canceled by the first term in (\ref{section2 equation 13}) were it expanded in a power series.  It, as well as $\Pi_4^-$, were calculated from the regulated expansion of the second term in (\ref{section2 equation 11}) using (\ref{section2 equation 12}).  

The quantity  $\Pi_4^-$   is obtained by adding all the fourth-order
terms in the expansion. It is the contribution of $H_-$  to the
photon-photon scattering graph. Its structure is
\begin{eqnarray}
\Pi_4^- = \Pi_4^{{scalar}} + \Pi_4^{\boldsymbol{\sigma} \cdot (\mathbf{{B}} - \mathbf{{E}}) }, 
\label{section2 equation 14}
\end{eqnarray}
where   $\Pi_4^{{scalar}}$ is the contribution to $\Pi_4^-$     neglecting the  $\boldsymbol{\sigma} \cdot  (\mathbf{{B}}-\mathbf{{E}})$
term in $V$. That is,   $\Pi_4^{{scalar}}$        is the contribution to $\Pi_4$   in scalar
QED$_4$  multiplied by $2$. The factor $2$, and not $4$, is due to the factor
of $\begin{matrix}\frac{1}{2} \end{matrix}$ in the definition (\ref{section2 equation 6}) of the spinor determinant. The
remainder in (\ref{section2 equation 14}) is the contribution to $\Pi_4^-$    from the  $\boldsymbol{\sigma} \cdot  (\mathbf{{B}}-\mathbf{{E}})$ term
in $V$. These two terms are
\begin{eqnarray}
\Pi_4^{{scalar}} &=& - \frac{1}{8\pi^2} \int \frac{d^4 k } {(2\pi)^4 } \frac{d^4 p}{ (2\pi)^4 } \frac{ d^4 q} {(2\pi)^4} \int_0^1 dz_1 z_1 \int_0^{1 - z_1}  dz_2 \int_0^{1 - z_1 - z_2 } dz_3 \nonumber \\
\nonumber\\
&\times&\left [ z_1 k^2  + z_2 p^2 + z_3  q^2 - (z_1 k + z_2 p + z_3 q)^2 + m^2 \right ]^{-2} \nonumber \\
&\times& \Big[ \begin{matrix}\frac{1}{4}\end{matrix}  (1 -2 z_1)^2 (1 - 2z_3)^2  \hat{F}_{\mu \nu} (p - k) \hat{F}_{\mu \nu} (k) \hat{F}_{\alpha \beta} (-q) \hat{F}_{\alpha \beta} (q - p)  \nonumber \\
&&+ \begin{matrix}\frac{1}{4} \end{matrix}  (1 - 2z_1 - 2z_2)^2 (1 - 2z_2 - 2z_3)^2 \hat{F}_{\mu \nu} (p - k) \hat{F}_{\mu \nu} (-q)  \hat{F}_{\alpha \beta} (k) \hat{F}_{\alpha \beta} (q - p) \nonumber \\
&&+  \begin{matrix}\frac{1}{4} \end{matrix}  (1 - 2z_2)^2 (1 - 2z_1 - 2z_2 - 2z_3)^2 \hat{F}_{\mu \nu} (p - k) \hat{F}_{\mu \nu} (q - p)  \hat{F}_{\alpha \beta} (k) \hat{F}_{\alpha \beta} (- q) \nonumber \\
&&- (1 -2z_1) (1 - 2z_2) (1 - 2z_3) (1 - 2z_1- 2z_2 - 2z_3) \hat{F}_{\alpha \beta} (p-k) \hat{F}_{\beta \gamma} (q - p) \hat{F}_{\gamma \delta} (-q) \hat{F}_{\delta \alpha} (k) \nonumber \\
&&+ (1 - 2z_1- 2z_2 - 2z_3) (1 - 2z_2) (1 - 2z_2 - 2z_3) ( 1- 2z_1 - 2z_2) \nonumber \\
&& \times \hat{F}_{\alpha \beta} (p-k) \hat{F}_{\beta \gamma} (-q) \hat{F}_{\gamma \delta} (k) \hat{F}_{\delta \alpha} (q - p) \nonumber \\
&&+ (1 - 2z_1 ) (1 - 2z_3) (1 - 2z_1- 2z_2) (1 - 2z_2 - 2z_3)  \hat{F}_{\alpha \beta} (p-k) \hat{F}_{\beta \gamma} (-q) \hat{F}_{\gamma \delta} (q - p) \hat{F}_{\delta \alpha} (k) \nonumber \\
&&+ \mbox{ three additional field strength terms.} \Big]. \nonumber \\ 
\label{section2 equation 15}
\end{eqnarray}
and
\begin{eqnarray}
\Pi_4^{\boldsymbol{\sigma} \cdot (\mathbf{{B}} - \mathbf{{E}}) } & = & - \frac{1}{8\pi^2} \int \frac{d^4k}{(2\pi)^4}  \frac{d^4 p }{(2\pi)^4} \frac{d^4 q} {(2\pi)^4} \int_0^1 dz_1 z_1 \int_0^{1 -z_1} dz_2 \int_0^{1 - z_1 -z_2} dz_3 \nonumber \\
& \times& \left [ z_1 k^2  + z_2 p^2 + z_3 q^2 - (z_1 k + z_2 p + z_3 q)^2 + m^2 \right ]^{-2} \nonumber \\
& \times& \left \{  \Big [  (\mathbf{\hat{B}} - \mathbf{\hat{E}}) (q - p) \cdot (\mathbf{\hat{B}} - \mathbf{\hat{E}}) (p - k) - \begin{matrix} \frac{1}{2} \end{matrix} \left [ (1 - 2z_1 - 2z_2 - 2z_3)^2 + (1 - 2z_2)^2 \right ] \right. \nonumber \\
&& \times  \hat{F}_{\mu \nu} (p- k) \hat{F}_{\mu \nu} (q - p) \Big ] (\mathbf{\hat{B}} - \mathbf{\hat{E}}) (k) \cdot (\mathbf{\hat{B}} - \mathbf{\hat{E}}) (-q) \nonumber \\
&+& \left [ (\mathbf{\hat{B}} - \mathbf{\hat{E}})  (k) \cdot (\mathbf{\hat{B}} - \mathbf{\hat{E}})  (p-k) - \begin{matrix}\frac{1}{2}\end{matrix} \left [ (1 - 2z_3)^2 + (1 - 2z_1)^2 \right ] \hat{F}_{\mu \nu} (p-k) \hat{F}_{\mu \nu} (k) \right ] \nonumber \\
&& \times (\mathbf{\hat{B}} - \mathbf{\hat{E}}) (-q) \cdot (\mathbf{\hat{B}} - \mathbf{\hat{E}} )(q - p) \nonumber \\
&-&  \Big[ (\mathbf{\hat{B}} - \mathbf{\hat{E}})  (-q) \cdot (\mathbf{\hat{B}} - \mathbf{\hat{E}})  (p-k) + \begin{matrix}\frac{1}{2}\end{matrix} \left [ (1 - 2z_1 - 2z_2)^2 + (1 - 2z_2 - 2z_3)^2 \right ] \nonumber \\
&& \times \hat{F}_{\mu \nu} (-q) \hat{F}_{\mu \nu} (p-k) \Big] (\mathbf{\hat{B}} - \mathbf{\hat{E}}) (k) \cdot (\mathbf{\hat{B}} - \mathbf{\hat{E}} )(q - p) \nonumber \\
& +&  \mbox{ three  additional field strength terms} \Big \} . 
\label{section2 equation 16}
\end{eqnarray}

In (\ref{section2 equation 15}) the additional terms are gauge invariant expressions such as
\[
\hat{F}_{\mu \nu} (-q)  \hat{F}_{\mu \nu} (q - p) q_{\alpha}  \hat{F}_{\alpha \beta} (p - k)  \hat{F}_{\beta \gamma} (k) q_{\gamma}, 
\label{section2 equation 17}
\]
and in (\ref{section2 equation 16}) they are of a similar form, such as
\[
(\mathbf{\hat{B}} - \mathbf{\hat{E}}) (-q) \cdot (\mathbf{\hat{B}} - \mathbf{\hat{E}}) (q - p) q_\alpha \hat{F}_{\alpha \beta} (p - k) \hat{F}_{\beta \gamma} (k) q_\gamma .  
\label{section2 equation 18}
\]
The full $\Pi_4$    graph has been calculated by Karplus and Neuman \cite{bib_23},
although the above results do not appear in their paper. Results
(\ref{section2 equation 15}) and (\ref{section2 equation 16}) require extensive, but straightforward, calculation.
As a check on  $\Pi_4^{{scalar}} $       we can specialize to the case of constant
$\mathbf{{B}}$ and $\mathbf{{E}}$. Using $F_{\alpha \beta}  F_{\beta \gamma}  F_{\gamma \delta}  F_{\delta \alpha}  = 2(\mathbf{{B}}^2  + \mathbf{{E}}^2  )^2 - 4(\mathbf{{E}}\cdot \mathbf{{B}})^2 , \Pi_4^{{scalar}}$
reduces to
\begin{eqnarray}
\Pi_4^{{scalar}} = - \frac{V}{720 \pi^4 m^4} \left ( \begin{matrix}\frac{7}{4} \end{matrix} (\mathbf{{B}}^2 + \mathbf{{E}}^2)^2 - (\mathbf{{B}} \cdot \mathbf{{E}})^2 \right ), 
\label{section2 equation 19}
\end{eqnarray}
where $V$ is a Euclidean volume cutoff. This is Weisskopf's \cite{bib_22} and
Schwinger's \cite{bib_14} constant field result for scalar QED$_4$'s fourth-order
effective Lagrangian continued to Euclidean space modulo a factor of $-2$.
The factor $2$ was discussed above. The minus sign arises from the
difference in statistics: we are calculating a contribution to the
scalar QED$_4$   effective action from the spinor effective action.

    Continuing with our discussion of (\ref{section2 equation 13}), $\Delta_-$  in the last trace
is the exact negative chirality propagator $\langle x | (H_-  + m^2)^{-1}  | y \rangle $ and
 $\Delta$  is the scalar propagator $\langle x | (P^2  + m^2  )^{-1}  |y \rangle $. The regulating
exponentials have been removed as the terms in the trace are
fifth order and higher, and so the implicit loop integral is
unambiguous. We leave the discussion of $\Delta_-$   to Secs. \hspace{-2mm} VI and VII.

The $A^2$ insertions clutter up the trace in the sense that in a
perturbative expansion of the trace they can be neglected. This is
because they form part of gauge invariant expressions whose
structure is already determined by the $AP + PA$ terms in $V$.
We do not see any justification for neglecting such terms in a
nonperturbative treatment of the trace, but nevertheless this
remark should be kept in mind.

    In order to discuss the $m=0$ limit of (\ref{section2 equation 13}) we must be more
specific about the background gauge field.


\section{Background Gauge Fields}

    QED determinants in constant field backgrounds have volume
divergences and so are not defined on non-compact manifolds.
Instead one considers the associated
effective Lagrangians, which do make sense. In the simplest case
of the Euclidean QED$_2$  determinant there is just a constant magnetic
field, and the volume divergence arises from the degeneracy of
the Landau levels. In constant-field QED$_4$, by making two rotations (a Lorentz boost plus a rotation in Minkowski space) the operator $(P-A)^2$ can be transformed into the sum of two two-dimensional harmonic oscillator Hamiltonians, leading to a degeneracy factor that grows as a four-volume \cite{bib_16}.  The lesson is that constant fields have too
much degeneracy to define the determinant on a non-compact manifold.

    We have found that $O(2) \times O(3)$ symmetric background fields allow
a satisfactory definition of the QED$_4$  determinant and that they
are sufficiently tractable to permit substantial analytic analysis.
Such fields were first explicitly considered in QED$_4$ by Adler
 \cite{bib_25,bib_26}. 
In this paper these fields take the form \cite{bib_16,bib_26,bib_17}.
\begin{equation}
A_\mu (x) = M_{\mu \nu} x_\nu  a (r^2),
\label{section3 equation 1}
\end{equation}
where $M_{\mu \nu}$  is chosen to be antiself-dual and is given by
\begin{equation}
M_{\mu \nu} = \left ( \begin{array}{cccc} &&&-1 \\ && 1 & \\ &-1 && \\ 1 &&& \end{array} \right ). 
\label{section3 equation 2}
\end{equation}
This field has an $O(2) \times O(3)$ invariance, subgroups present in the
reduction of $O(4)$ to $O(3)\times O(3)$. It is further assumed that $a(r^2 )$
is smooth, well-behaved at the origin, and satisfies
\begin{equation}
a (r^2) = \frac{\nu}{r^2}, \; \; r > R, 
\label{section3 equation 3}
\end{equation}
where $\nu$   is a dimensionless constant. Without loss of generality
assume $\nu >0$.

    The orbital angular momentum operators of the first and second
$O(3)$ subgroups of $O(4)$ are
\begin{eqnarray}
L_k^{(1)} = \begin{matrix}\frac{1}{2}\end{matrix} i \left(x_0 \frac{\partial}{\partial x_k} - x_k \frac{\partial}{\partial x_0} - \epsilon_{klm}x_l \frac{\partial}{\partial x_m} \right),\nonumber \\
\nonumber\\
L_k^{(2)}  = \begin{matrix}\frac{1}{2}\end{matrix} i \left(x_k \frac{\partial}{\partial x_0} - x_0 \frac{\partial}{\partial x_k} - \epsilon_{klm}x_l \frac{\partial}{\partial x_m} \right), 
\label{section3 equation 4}
\end{eqnarray}
which satisfy
\begin{eqnarray}
\left[ L_i^{(p)} , L_j^{(q)} \right] = \delta_{pq} i \epsilon_{ijk} L_k^{(p)}, p, q, = 1, 2.
\label{section3 equation 5}
\end{eqnarray}
The spin angular momentum operators in the representation (\ref{section2 equation 7}) are
\begin{eqnarray}
S_k^{(1)} = \begin{matrix}\frac{1}{2}\end{matrix} \left ( \begin{array}{cc} \sigma_k & 0 \\ 0 & 0 \end{array} \right ) \mbox{ ,  }
S_k^{(2)} = \begin{matrix}\frac{1}{2}\end{matrix} \left ( \begin{array}{cc} 0 & 0 \\ 0 & \sigma_k \end{array} \right ). 
\label{section3 equation 6}
\end{eqnarray}
The total angular momentum operator relative to the second subgroup,
\begin{eqnarray}
J_k^{(2)} = L_k^{(2)} + S_k^{(2)}, 
\label{section3 equation 7}
\end{eqnarray}
commutes with $\displaystyle{\not}A$:
\begin{eqnarray}
\left[ J_k^{(2)}, \displaystyle{\not}A \right] = 0, k = 1, 2, 3, 
\label{section3 equation 8}
\end{eqnarray}
while $\displaystyle{\not}A$ is invariant only with respect to rotations about the third
axis of the first subgroup:
\begin{eqnarray}
\left[ J_3^{(1)}, \displaystyle{\not}A \right] = 0.
\label{section3 equation 9}
\end{eqnarray}
We adopt the conventions of \cite{bib_19} 
for the four-dimensional rotation
matrices $D_{m_1 m_2}^l$:
\begin{eqnarray}
L^{(1)}\cdot L^{(1)} D_{m_1 m_2}^l (x) &=& l ( l + 1) D_{m_1 m_2}^l (x), \nonumber \\
L_3^{(1)} D_{m_1 m_2}^l (x) &=& m_1 D_{m_1 m_2}^l (x), \nonumber \\
L_3^{(2)} D_{m_1 m_2}^l (x) &=& m_2 D_{m_1 m_2}^l (x) .
\label{section3 equation 10}
\end{eqnarray}
The $D_{m_1 m_2}^l$   are normalized so that
\begin{eqnarray}
\int  d \Omega_4 D_{m_1 m_2}^{l_1 *} (x) D_{m_3 m_4}^{l_2} (x)  = \frac{2\pi^2}{2 l_1 +1} \delta_{l_1 l_2} \delta_{m_1 m_3} \delta_{m_2 m_4} (r^2)^{2 l_1},
\label{section3 equation 11}
\end{eqnarray}
where  $\Omega_4$   is the surface element in four dimensions. Some properties
of these matrices appear in Appendix A.

    Following \cite{bib_19} 
we construct eigenstates of $J^{(1)}\cdot J^{(1)}, J_3^{(1)}$  (eigenvalues
$j\pm\frac{1}{2}  , M)$ and $J^{(2)}\cdot  J^{(2)}, J_3^{(2)}$  (eigenvalues $j, m$). In the positive chirality
sector these are
\begin{eqnarray}
\varphi_{j, m}^{j\pm\frac{1}{2}, M}(x) = 
\left (
\begin{array}{c}
\mp (j \pm M + \frac{1}{2})^\frac{1}{2}D_{M-\frac{1}{2}, m}^j (x) \\
 (j \mp M + \begin{matrix}\frac{1}{2}\end{matrix})^\frac{1}{2}D_{M+\frac{1}{2}, m}^j (x) \\
0\\
0
\end{array} \right ),
\label{section3 equation 12}
\end{eqnarray}
and in the negative chirality sector they are
\begin{eqnarray}
\psi_{j, m}^{j+\frac{1}{2}, M}(x) = 
\left (
\begin{array}{c}
0\\
0\\
- (j - m + 1)^\frac{1}{2}D_{M, m-\frac{1}{2}}^{j+\frac{1}{2}} (x) \\
 (j + m + 1 )^\frac{1}{2}D_{M, m+\frac{1}{2}}^{j+\frac{1}{2}} (x) 
\end{array} \right ),
\label{section3 equation 13}
\end{eqnarray}
\begin{eqnarray}
\psi_{j, m}^{j-\frac{1}{2}, M}(x) = 
\left (
\begin{array}{c}
0\\
0\\
(j + m )^\frac{1}{2}D_{M, m-\frac{1}{2}}^{j-\frac{1}{2}} (x) \\
 (j - m )^\frac{1}{2}D_{M, m+\frac{1}{2}}^{j-\frac{1}{2}} (x) 
\end{array} \right ).
\label{section3 equation 14}
\end{eqnarray}
Due to (\ref{section3 equation 8}) and (\ref{section3 equation 9}), eigenstates of $\displaystyle{\not}D = \displaystyle{\not}P - \displaystyle{\not}A$ are of the form \cite{bib_19}
\begin{eqnarray}
\psi_{EjMm}^+ (x) = F (r^2)\varphi_{j,m}^{j-\frac{1}{2}, M} (x) + G(r^2) \varphi_{j,m}^{j + \frac{1}{2}, M} (x),
\label{section3 equation 15}
\end{eqnarray}
\begin{eqnarray}
\psi_{EjMm}^- (x) = f (r^2)\psi_{j,m}^{j-\frac{1}{2}, M} (x) + g(r^2) \psi_{j,m}^{j + \frac{1}{2}, M} (x),
\label{section3 equation 16}
\end{eqnarray}
where the superscripts on $\psi_{EjMm}^{\pm}$      denote chirality and $E$ is the energy
eigenvalue.   In the following we will write $\psi_{EjMm}^{\pm}$ as two-component spinors.  

From   $^*\hspace{-1mm}F_{\mu \nu}   = \frac{1}{2}\epsilon^{\mu \nu\alpha \beta}F_{\alpha \beta}    $  and (\ref{section3 equation 1}) it follows that
\begin{eqnarray}
^*F_{\mu \nu} F_{\mu \nu} = -16a^2 - 16r^2 a a',
\label{section3 equation 17}
\end{eqnarray}
and
\begin{eqnarray}
F_{\mu \nu} F_{\mu \nu} = 8r^4a^{'2} - ^*\hspace{-1mm}F_{\mu \nu} F_{\mu \nu},
\label{section3 equation 18}
\end{eqnarray}
where the prime denotes differentiation with respect to $r^2$. 
From (\ref{section3 equation 3}) and (\ref{section3 equation 17}) the chiral anomaly is
\begin{eqnarray}
-\frac{1}{16 \pi^2} \int d^4x\hspace{1mm} ^*\hspace{-1mm}F_{\mu \nu} F_{\mu \nu} = \frac{\nu^2}{2},
\label{section3 equation 19}
\end{eqnarray}
provided $\underset{r \rightarrow 0}{\lim}\; r^2 a = 0$.  
Note, as expected, that $F_{\mu \nu}$  is not square-integrable. But this does not
matter as far as the remainder $\mathcal{R}$ in (\ref{section2 equation 10}) is concerned. Recall that
it is only required that $A_\mu \in \bigcap\limits_{n>4}L^n (\mathbb{R}^4)$, which it does here. Furthermore,
because we have chosen on-shell charge renormalization 
the $1/k^2$ behavior of $\hat{F}_{\mu\nu}$ for small $k$ 
in the first term on the right-hand side of (\ref{section2 equation 6})
is regulated by the vanishing logarithm as $k\rightarrow 0$.
So everything in (\ref{section2 equation 6}) is finite.

If one wishes to deal with a negative anomaly then instead of
choosing $M_{\mu \nu}$   in (\ref{section3 equation 1}) antiself-dual, require it to be self-dual.
For example, let

\begin{equation}
M_{\mu \nu} \rightarrow N_{\mu \nu} = \left ( 
\begin{array}{cccc}
&&&1\\
&&1&\\
&-1&&\\
-1&&&\\
\end{array}
\right ).
\label{section3 equation 20}
\end{equation}


\section{ZERO MASS LIMIT OF $\mathcal{R}$   : PRELIMINARIES}

    We will now discuss in a preliminary way the limit of (\ref{section2 equation 13}) as
$m \rightarrow 0$. Consider the first term. A working definition of the chiral
anomaly for $\displaystyle{\not}{D}$ on non-compact manifolds is \cite{bib_15}.
\begin{eqnarray}
\lim_{m \rightarrow 0} m^2 \mbox{Tr} \left[ (H_+ + m^2)^{-1} - (H_- + m^2)^{-1} \right ] = -\frac{1}{16\pi^2} \int d^4x {^{\phantom{1}*}} F_{\mu \nu} F_{\mu \nu}.
\label{newsection4 equation 1}
\end{eqnarray}
Because the manifold is a non-compact Euclidean one the right-hand
side of (\ref{newsection4 equation 1}) need not be the difference between numbers $n_+  - n_-$  of
positive and negative chirality $L^2$   zero modes. The remainder, if
any, is related to the zero-energy phase shifts associated with $H_\pm$
\cite{bib_15}. More will be said about this at the end of Sec. V. If the
remaining terms in (\ref{section2 equation 13}) vanish in the $m=0$ limit then (\ref{section3 equation 19}) and (\ref{newsection4 equation 1})
indicate that $\mathcal{R}$   in (\ref{section2 equation 10}) behaves as
\begin{eqnarray}
\mathcal{R} \underset{m\rightarrow 0 } {\sim} \frac{\nu^2}{4} \ln m^2 + \mbox{less singular in } m^2.  
\label{newsection4 equation 2}
\end{eqnarray}
Thus, $\mathcal{R}$    would become negative as $ m\rightarrow 0$.

    A necessary condition for the vanishing of the remaining terms
is that there be no $L^2 $ zero modes in the negative chirality sector.
It will be shown in Sec. V that this is true for our choice of
gauge fields. Otherwise, $\Delta_-$   in (\ref{section2 equation 13}) would develop a simple pole at
$m  = 0$ and (\ref{newsection4 equation 2}) would contain more terms varying as $\ln m^2$ for $m \rightarrow 0$. But this is not a sufficient
condition for the remaining terms in (\ref{section2 equation 13}) to vanish at $m=0$.

    One can see already from the second term in (\ref{section2 equation 13}) some of the
subtleties involved. If $\mathbf{{B}}(x)$ and $\mathbf{{E}}(x)$ fall off as $1/r^2$  , as our
fields do, without any particular symmetry constraint then their
Fourier transforms will be such that $\mathbf{\hat{B}}(k)$, $\mathbf{\hat{E}}(k)$ behave as $1/k^2$  as
$k \rightarrow 0$. In this case the integral will have an infrared divergence
even when $m \not= 0$. But this does not happen due to the $O(2) \times O(3)$
symmetry of the gauge fields.

    To see this define
\begin{eqnarray}
\hat{F}_{\mu \nu}^> (k) = \int_{|x| > R} d^4 x e^{- ikx} F_{\mu \nu} (x), 
\label{newsection4 equation 3}
\end{eqnarray}
with
\begin{eqnarray}
F_{\mu \nu} \underset{|x| > R} {=}  - \frac{2 \nu }{r^2} \left ( M_{\mu \nu} + \frac{x_\mu M_{\nu \alpha} x_\alpha - x_\nu M_{\mu \alpha} x_\alpha }{r^2} \right ).
\label{newsection4 equation 4}
\end{eqnarray}
Then
\begin{eqnarray}
\hat{F}_{\mu \nu}^> (k) = 
\frac{8\pi^2 \nu}{k^2} \left [ 
M_{\mu \nu} J_2 (kR) + \frac{M_{\nu \alpha} k_\alpha k_\mu - M_{\mu \alpha} k_\alpha k_\nu} { k^2} (J_0 (kR) + 2J_2 (kR) )  
\right ], 
\label{newsection4 equation 5}
\end{eqnarray}
and
\begin{eqnarray}
\mathbf{\hat{B}}^> (k) \cdot \mathbf{\hat{E}}^> (-k) = - \frac{(8 \pi^2 \nu)^2}{k^4} ( J_0 (kR) J_2 (kR) + J_2^2 (kR) ) . 
\label{newsection4 equation 6}
\end{eqnarray}
Thus, $\mathbf{\hat{B}}^> (k)\cdot  \mathbf{\hat{E}}^> (-k)$  behaves as $R^2 /k^2$  instead of $1/k^4$  as $k\rightarrow 0$. For large
$k$, $\hat{F}_{\mu \nu}^<  (k)$, calculated from an integral like (\ref{newsection4 equation 3}) but with $|x|<R$,
behaves as $\sin(kR-3 \pi /4)/k^{5/2}$    for any reasonable behavior of $a(r^2 )$
near $r=0$, such as a  $\underset{r \rightarrow 0}{\sim}   \mbox{Cr}^{\beta}$  with  $\beta  > -\frac{1}{2}$ or $-\frac{1}{3}$ as required in
Sec. VIII. Therefore, the integral in (\ref{section2 equation 13}) is absolutely convergent
in the ultraviolet and its small-mass limit varies as $(\ln m^2)^2$, allowing us to conclude that the second term
in (\ref{section2 equation 13}) vanishes in the limit $m=0$.

    Now consider the third term in (\ref{section2 equation 13}), $m^2 \partial \Pi_4^- /\partial m^2$. Referring to
(17), (18), (19) and (\ref{newsection4 equation 5}), simple power counting of momenta
suggests that the integrals defining $\Pi_4^{{scalar}}$           and  $\Pi_4^{\boldsymbol{\sigma}(\mathbf{B} - \mathbf{E}) }$       have
a logarithmic mass singularity of the form $(\ln m^2 )^n$  with $n \geq 1$. If so,
then the $m=0$ limit of $m^2\partial \Pi_4^- /\partial m^2$          would be nonvanishing,  thereby
falsifying (\ref{newsection4 equation 2}). It is encouraging that there is no immediate
infrared divergence for $m=0$ that has to be canceled by the
symmetry of $F_{\mu\nu}$   , as in the second-order term of (\ref{section2 equation 13}). The confluence
of singularities in $\mathbf{\hat{E}}$ and $\mathbf{\hat{B}}$ is no longer present in fourth order.
Nor are they present in higher orders due to the result cited in
Sec. II that $\det_5$   is well-defined for any $A_\mu \in \underset{n > 4}{\cap}     L^n ({\mathbb R^4}  )$, which
includes our fields.

    The fact that power counting does not ensure finiteness in the
$m=0$ limit of $\Pi_4^-$    indicates that the symmetry properties of $\hat{F}_{\mu \nu}  (k)$
will be required to give a finite limit to the individual terms in
(18) and (19). Because of this reliance on symmetry, theorems on
mass singularities of Feynman amplitudes known to the author are
inapplicable here. The analysis required to give a definitive
answer one way or another is beyond the scope of this paper. All
we are able to do here is to present evidence for a finite
limit of $\Pi_4^-$    as $ m \rightarrow 0$.
We note that Adler's stereographic mapping to the surface
    of a 5-dimensional unit hypersphere \cite{bib_25},\cite{bib_26} cannot help
    here due to the slow $1/r$ falloff of the vector potential.

    Consider, for example, the fourth term in (18). Let $p \rightarrow p+k$,
$q \rightarrow -q$ so that the chain of field strengths is put into the form
\begin{eqnarray}
\hat{F}_{\alpha \beta} (p) \hat{F}_{\beta \gamma} (-k - p -q) \hat{F}_{\gamma \delta} (q) \hat{F}_{\delta \alpha} (k) &=&  \begin{matrix}\frac{1}{4}\end{matrix} \hat{F}_{\alpha \beta} (p) \hat{F}_{\alpha \beta} (-k-p-q) \hat{F}_{\mu \nu} (q) \hat{F}_{\mu \nu} (k) \nonumber \\
&+& \begin{matrix}\frac{1}{4}\end{matrix} \hat{F}_{\alpha \beta} (p) \hat{F}_{\alpha \beta} (k) \hat{F}_{\mu \nu} (-k-p-q) \hat{F}_{\mu \nu} (q) \nonumber \\
&-& \left [ \mathbf{\hat{B}} (p) \cdot \mathbf{\hat{E}} (q) + \mathbf{\hat{B}} (q) \cdot \mathbf{\hat{E}} (p) \right ]  \nonumber \\
& & \times \left [  \mathbf{\hat{B}} (-k-p-q) \cdot \mathbf{\hat{E}} (k) + \mathbf{\hat{B}} (k) \cdot \mathbf{\hat{E}} (-k-p-q) \right ].\nonumber \\ 
\label{newsection4 equation 7}
\end{eqnarray}
The $1/k^2$  behavior of $\hat{F}_{\mu \nu}  (k)$ arises from the $J_0 (kR)$ term in (\ref{newsection4 equation 5}).
Fixing on the most singular terms, the first term on the
right-hand side of (\ref{newsection4 equation 7}) contributes
\begin{eqnarray}
\hat{F}_{\mu \nu} (q) \hat{F}_{\mu \nu} (k) \underset{k, q \rightarrow 0} { =} \frac{128 \pi^4 \nu^2}{k^4 q^4} 
\left [ 
(q \cdot k)^2  - (q_0 k_3 - q_1 k_2 + q_2 k_1 - q_3 k_0)^2 
\right ].
\label{newsection4 equation 8}
\end{eqnarray}
To isolate the leading singularity when $k,q\rightarrow 0$ we neglect the
denominator in the fourth term in (18) when integrating over the
angles defining $q_\mu$  . Using
\begin{eqnarray}
\int d\Omega_q (k \cdot q)^2 &=& \frac{\pi^2}{2} k^2 q^2, \nonumber \\
\int d\Omega_q (q_0 k_3 - q_1 k_2 + q_2 k_1 - q_3 k_0)^2 &=& \frac{\pi^2}{2} k^2 q^2, 
\label{newsection4 equation 9}
\end{eqnarray}
we see that the leading singularity for small $k$ and $q$ cancels. The
first term on the right-hand side of (\ref{newsection4 equation 7}) also contains the term
$F_{\alpha \beta}  (p)F_{\alpha \beta}  (-k-p-q)$. The case $k,p,q \rightarrow 0$  with $p<<k,q$ reduces to the
case just considered when the angles defining $p_\mu$  are integrated
over. The same conclusions follow for the small $k$ and $p$ behavior of
the second term in (\ref{newsection4 equation 7}) as well as the case $k,p,q \rightarrow 0$ with $q<<p,k$.

    Finally, consider the third term on the right-hand side of
(\ref{newsection4 equation 7}). Referring to the result (\ref{newsection4 equation 6}), the singularity in
$\mathbf{\hat{B}}(-k-p-q) \cdot \mathbf{\hat{E}}(k)$ and $\mathbf{\hat{B}}(k) \cdot \mathbf{\hat{E}}(-k-p-q)$ is $R^2 /k^2 $ at $p,q=0$ and not $1/k^4$.

    The gauge invariant expressions on the right-hand side of (\ref{newsection4 equation 7})
occur in all the terms in (18) and (19), and so the above
cancellations occur there too. There are three additional field
strength terms in the integrands of (18) and (19) and these must
also be considered.

    Given that the second and fourth-order terms in (16) vanish
at $m=0 $ then presumedly so will all higher order terms  $m^2 \partial \Pi_6^- /\partial m^2, \dots $  
generated by expanding $\Delta_-$   in (16). Since the zero modes reside in
the positive chirality propagator $\Delta_+$   this expansion may have some
justification. However, the scattering states extend down to zero
energy, and these may result in nonperturbative mass singularities
induced by $\Delta_-$. This will be examined in Secs. VI and VII.

\section{Zero Modes}
In the representation (\ref{section2 equation 7}) $\displaystyle{\not}D$ has the supersymmetric structure
\begin{eqnarray}
\displaystyle{\not}D = \left (
\begin{array}{cc}
0 & D \\
-D^\dagger & 0 \\
\end{array}
\right),
\label{section4 equation 1}
\end{eqnarray}
and hence positive chirality zero modes are square-integrable
solutions of
\begin{eqnarray}
D^\dagger \psi^+ = 0,
\label{section4 equation 2}
\end{eqnarray}
where all subscripts on $\psi^+$
   have been dropped. From (\ref{section3 equation 12}) and (\ref{section3 equation 15})
\begin{eqnarray}
\psi^+(x) = \left (
\begin{array}{c}
\left[ (j - M+\begin{matrix}\frac{1}{2}\end{matrix})^\frac{1}{2}F - (j+M+\begin{matrix}\frac{1}{2}\end{matrix})^\frac{1}{2} G \right] D_{M-\frac{1}{2}, m}^j (x) \\
\\
\left[ (j + M+\begin{matrix}\frac{1}{2}\end{matrix})^\frac{1}{2}F + (j-M+\begin{matrix}\frac{1}{2}\end{matrix})^\frac{1}{2} G \right] D_{M+\frac{1}{2}, m}^j (x) \\
\end{array}
\right).
\label{section4 equation 3}
\end{eqnarray}
By (\ref{section3 equation 11}), $\psi^+ \in L^2$  provided
\begin{equation}
	\int_0^\infty dr r^{4j + 3} ( F^2 + G^2) < \infty.
\label{section4 equation 4}
\end{equation}
Inserting (\ref{section4 equation 3}) in (\ref{section4 equation 2}) results in
\begin{equation}
G' + \frac{a}{2j + 1} \left ( \sqrt { (j+\begin{matrix}\frac{1}{2}\end{matrix} )^2 -M^2 }F - MG\right) = 0,
\label{section4 equation 5}
\end{equation}
\begin{equation}
r^2 F' + (2j + 1)F + \frac{ar^2}{2j + 1} \left(M F + \sqrt{ (j+\begin{matrix}\frac{1}{2}\end{matrix})^2 -M^2} G \right)= 0.
\label{section4 equation 6}
\end{equation}
Here $j = 0, \frac{1}{2}, \dots $    and $-j-\frac{1}{2}  \leq M \leq j +\frac{1}{2}$. Equations (\ref{section4 equation 5}) and (\ref{section4 equation 6})
appear in  \cite{bib_18, bib_19} 
in a different notation, although the authors are
considering an entirely different problem. There are three cases to
consider.

    Case 1: $M = -j-\frac{1}{2}$. Then
\begin{equation}
\psi^+ = \sqrt{2j + 1} D_{-j, m}^j (x) G(r^2) \left ( \begin{array}{c} 0 \\ 1 \end{array} \right ),
\label{section4 equation 7}
\end{equation}
with
\begin{equation}
\frac{dG}{dr^2} = -\frac{a}{2} G,
\label{section4 equation 8}
\end{equation}
and so
\begin{equation}
G(r^2) = G(r_0^2) e^{-\frac{1}{2} \int_{r_0^2}^{r^2} ds a (s) }.
\label{section4 equation 9}
\end{equation}
Since $a = \nu/r^2$  for $r>R$, $\psi^+ \in L^2$  for $j = 0,\frac{1}{2},...    , j_{max}    $, where $j_{max}$   is the
largest value of $j$ for which $\nu  > 2j + 2$ is satisfied.

    Case 2: $M = j +  \frac{1}{2}$ . Then
\begin{eqnarray}
\psi^+ = -\sqrt{2j + 1} D_{j,m}^j (x) G(r^2) \left( \begin{array}{c} 1 \\ 0 \end{array} \right ),
\label{section4 equation 10}
\end{eqnarray}
and
\begin{eqnarray}
\frac{dG}{dr^2} = \frac{a}{2} G. 
\label{section4 equation 11}
\end{eqnarray}
Based on case 1 it is clear that $G \displaystyle{\not}{\in}  L^2$  for any  $\nu > 0$.

    Case 3: $|M|<j + \frac{1}{2}$ . We claim that there are no $L^2$  zero modes in
this case. To show this let $z = r^2$  ,
\begin{eqnarray}
\triangle = (2j + 1) F, 
\label{section4 equation 12}
\end{eqnarray}
\begin{eqnarray}
\Gamma = - 2MF - 2 \sqrt{ (j+\begin{matrix}\frac{1}{2}\end{matrix})^2 - M^2 } G,
\label{section4 equation 13}
\end{eqnarray}
in (\ref{section4 equation 5}) and (\ref{section4 equation 6}). Then these become
\begin{eqnarray}
z \frac{d \triangle}{dz} + (2 j + 1) \triangle = \begin{matrix}\frac{1}{2}\end{matrix} z a \Gamma,
\label{section4 equation 14}
\end{eqnarray}
\begin{eqnarray}
z \frac{d \Gamma}{dz} = (2M + \begin{matrix}\frac{1}{2}\end{matrix} z a ) \triangle.
\label{section4 equation 15}
\end{eqnarray}
Assume $a$ has a power series expansion about $z = 0$ and let
\begin{eqnarray}
a & = & \sum_0 a_n z^n, \nonumber \\
\triangle& = &\sum_0 b_n z^n, \nonumber \\
\Gamma& =& \sum_0 c_n z^n.
\label{section4 equation 16}
\end{eqnarray}
Then $\triangle$    and $\Gamma$   have the expansions
\begin{eqnarray}
\triangle = c_0 \left ( 
\frac{a_0}{4 (j+1)} z + 
\left ( \frac{a_1}{2(2j+3) } + \frac{M a_0^2}{4(j + 1)(2 j +3) } \right ) z^2 + O (z^3)
\right ),
\label{section4 equation 17}
\end{eqnarray}
\begin{eqnarray}
\Gamma = c_0 \left ( 
1 + \frac{M a_0}{2 (j+1)} z + 
\left ( \frac{M a_1}{ 2(2j+3) }+ \frac{M^2 a_0^2}{4(j + 1)(2 j +3) } + \frac{a_0^2}{16 (j+1) } \right ) z^2 + O (z^3)
\right ).\nonumber \\
\label{section4 equation 18}
\end{eqnarray}

It will now be shown that the solution  $\triangle$  , $\Gamma$   that is finite
at $r = 0$ does not converge fast enough to make  $\psi^+ \in L^2$  for $|M|<j + \frac{1}{2}$
with $\psi^+$   given by (\ref{section4 equation 3}). Let $t = \ln z$ and
\begin{eqnarray}
\Gamma &=& \gamma e^{-(j + \frac{1}{2} ) t}, \nonumber \\
\triangle &=& \delta e^{-(j + \frac{1}{2} ) t}, \nonumber \\
a &=& \alpha e^{-t}. 
\label{section4 equation 19}
\end{eqnarray}
Then (\ref{section4 equation 14}) and (\ref{section4 equation 15}) become
\begin{eqnarray}
\frac{d\delta}{dt} + ( j + \begin{matrix}\frac{1}{2}\end{matrix} ) \delta = \begin{matrix} \begin{matrix}\frac{1}{2}\end{matrix}\end{matrix} \alpha \gamma,
\label{section4 equation 20}
\end{eqnarray}
\begin{eqnarray}
\frac{d \gamma}{dt} - (j + \begin{matrix}\frac{1}{2}\end{matrix}) \gamma = (2 M + \begin{matrix}\frac{1}{2}\end{matrix} \alpha ) \delta.
\label{section4 equation 21}
\end{eqnarray}
These are the same equations appearing in Eq. (5.24) of \cite{bib_18}. 
Following
their analysis, multiply (\ref{section4 equation 20}) by $\delta$  , (\ref{section4 equation 21}) by $\gamma$  and subtract:
\begin{eqnarray}
\begin{matrix}\frac{1}{2}\end{matrix} \frac{d}{dt} (\gamma^2 - \delta^2) = (j + \begin{matrix}\frac{1}{2}\end{matrix}) ( \gamma^2 + \delta^2) + 2M\gamma \delta. 
\label{section4 equation 22}
\end{eqnarray}
Since   $\gamma=  r^{2j+1}\Gamma$, $\delta  = r^{2j+1} \triangle$     and  $\Gamma$, $\triangle$ are finite at $r=0$, $\gamma$  and $\delta$
vanish at $r=0$. From (\ref{section4 equation 4}), if $\psi^+ \in L^2 $  then $ F$, $G \sim  r^{-2j - 2 - \epsilon}$, $\epsilon    >0$ and hence
$\gamma, \delta \sim r^{-1-\epsilon}$  for $r\rightarrow \infty$. Integrating (\ref{section4 equation 22}) therefore gives
\begin{eqnarray}
\int_0^\infty \frac{dr}{r} \left[ (j + \begin{matrix}\frac{1}{2}\end{matrix}) (\gamma^2 + \delta^2) + 2 M \gamma \delta \right] = 0. 
\label{section4 equation 23}
\end{eqnarray}
Since $|M|< j + \frac{1}{2}$,    (\ref{section4 equation 23}) is impossible for real $e$. Hence the
assumption that $\psi^+ \in L^2$   for $|M|< j + \frac{1}{2}$  is false.

    We now turn to the negative chirality sector. From (\ref{section3 equation 13}), (\ref{section3 equation 14}) and
(\ref{section3 equation 16}),
\begin{equation}
\psi^- (x) = \left ( 
\begin{array}{c}
(j+m)^\frac{1}{2} D_{M,m-\frac{1}{2}}^{j - \frac{1}{2}} (x) \\
(j-m)^\frac{1}{2} D_{M,m+\frac{1}{2}}^{j - \frac{1}{2}} (x) 
\end{array}
\right ) f(r^2) + 
\left (
\begin{array}{c}
- ( j-m+1)^{\frac{1}{2}} D_{M, m - \frac{1}{2}}^{j + \frac{1}{2}} (x) \\
 ( j+m+1)^{\frac{1}{2}} D_{M, m + \frac{1}{2}}^{j + \frac{1}{2}} (x) 
\end{array}
\right ) g(r^2),
\label{section4 equation 24}
\end{equation}
and $\psi^- \in L^2$  provided
\begin{equation}
\int_0^\infty dr r^{4j+1} \left[ f^2 + (r^2g)^2 \right ] < \infty. 
\label{section4 equation 25}
\end{equation}
From (\ref{section4 equation 1}) negative chirality zero modes are $L^2$ solutions of
\begin{equation}
D\psi^- = 0.
\label{section4 equation 26}
\end{equation}
Substitution of (\ref{section4 equation 24}) in (\ref{section4 equation 26}) results in
\begin{equation}
2 \sqrt{ j - M + \begin{matrix}\frac{1}{2}\end{matrix} } f' - \sqrt{ j + M + \begin{matrix}\frac{1}{2}\end{matrix} } \left( 2 r^2 g' + 4 (j + 1) g \right) + 
\sqrt{ j - M + \begin{matrix}\frac{1}{2}\end{matrix} } a f - \sqrt{j + M + \begin{matrix}\frac{1}{2}\end{matrix}}  r^2 a g = 0,
\label{section4 equation 27}
\end{equation}
\begin{equation}
2 \sqrt{ j + M + \begin{matrix}\frac{1}{2}\end{matrix} } f' + \sqrt{ j - M + \begin{matrix}\frac{1}{2}\end{matrix} } \left( 2 r^2 g' + 4 (j + 1) g \right) - 
\sqrt{ j + M + \begin{matrix}\frac{1}{2}\end{matrix} } a f - \sqrt{j - M + \begin{matrix}\frac{1}{2}\end{matrix}}  r^2 a g = 0. 
\label{section4 equation 28}
\end{equation}
There are again three cases.

    Case 1: $M = -j -\frac{1}{2}$  . From (\ref{section4 equation 24}),
\begin{equation}
\psi^- (x) = \left ( 
\begin{array}{c}
-(j - m+1)^{\frac{1}{2}}D_{-j - \frac{1}{2}, m - \frac{1}{2}}^{j + \frac{1}{2}} (x)\\
(j +  m+1)^{\frac{1}{2}}D_{-j - \frac{1}{2}, m + \frac{1}{2}}^{j + \frac{1}{2}} (x)
\end{array}
\right ) g(r^2).
\label{section4 equation 29}
\end{equation}
From (\ref{section4 equation 28})
\begin{equation}
2 r^2g' + 4(j+1)g - r^2 a g = 0,
\label{section4 equation 30}
\end{equation}
whose solution by inspection is
\begin{equation}
g(r^2) = g(r_0^2) \left( \frac{r}{r_0} \right) ^{-4 j - 4} e^{\frac{1}{2} \int_{r_0^2}^{r^2} ds a (s)}. 
\label{section4 equation 31}
\end{equation}
By (\ref{section4 equation 25})    $\psi^- \in L^2 $  only if
\begin{equation}
\int_0^\infty dr r^{4j + 5} g^2 < \infty,
\label{section4 equation 32}
\end{equation}
and therefore $g$ is too singular at $r = 0$ to be in $L^2$ .

    Case 2: $M = j + \begin{matrix}\frac{1}{2}\end{matrix}$ . From (\ref{section4 equation 24}),
\begin{eqnarray}
\psi^- = \left ( 
\begin{array}{c}
- (j - m + 1)^{\frac{1}{2}} D_{j+\frac{1}{2},m - \frac{1}{2}}^ {j + \frac{1}{2}} (x) \\
(j + m + 1)^{\frac{1}{2}} D_{j+\frac{1}{2},m + \frac{1}{2}}^{j + \frac{1}{2}} (x) 
\end{array}
\right ) g (r^2). 
\label{section4 equation 33}
\end{eqnarray}
and from (\ref{section4 equation 27}),
\begin{eqnarray}
2r^2 g' + 4 (j+1)g + r^2 a g  = 0.
\label{section4 equation 34}
\end{eqnarray}
As in case 1 $g$ is too singular at $r = 0$ to be in $L^2$  .

    Case 3: $|M| < j +  \frac{1}{2}$. We will demonstrate that $\psi^- \displaystyle{\not}{\in}   L^2$ .
Let $z = r^2$  ,
\begin{eqnarray}
\Gamma = \sqrt{( j + M + \begin{matrix}\frac{1}{2}\end{matrix} ) } ( f + r^2 g) + \sqrt{(j - M + \begin{matrix}\frac{1}{2}\end{matrix}) } (r^2g - f), 
\label{section4 equation 35}
\end{eqnarray}
\begin{eqnarray}
\triangle = \sqrt{(j + M + \begin{matrix}\frac{1}{2}\end{matrix})} ( f - r^2 g) + \sqrt{( j - M + \begin{matrix}\frac{1}{2}\end{matrix}) } (f + r^2g).
\label{section4 equation 36}
\end{eqnarray}
Then (\ref{section4 equation 27}), (\ref{section4 equation 28}) become
\begin{eqnarray}
z \frac{d\triangle}{dz} + \left( j + \begin{matrix}\frac{1}{2}\end{matrix} - \sqrt{ (j + \begin{matrix}\frac{1}{2}\end{matrix})^2 - M^2} \right ) \triangle = (M + \begin{matrix}\frac{1}{2}\end{matrix} z a ) \Gamma,
\label{section4 equation 37}
\end{eqnarray}
\begin{eqnarray}
z \frac{d\Gamma}{dz} + \left ( j + \begin{matrix}\frac{1}{2}\end{matrix} + \sqrt{ (j + \begin{matrix}\frac{1}{2}\end{matrix})^2 - M^2} \right ) \Gamma = (M + \begin{matrix}\frac{1}{2}\end{matrix} z a ) \triangle.
\label{section4 equation 38}
\end{eqnarray}
Making the expansions (\ref{section4 equation 16}) gives for $M \not= 0$,
\begin{eqnarray}
\triangle = b_0 \left ( 
1 + \frac{ 2M^2 + j + \begin{matrix}\frac{1}{2}\end{matrix} - \sqrt{ (j+ \begin{matrix}\frac{1}{2}\end{matrix})^2 - M^2 } }{4M(j+1)} a_0 z + O(z^2) 
\right),
\label{section4 equation 39}
\end{eqnarray}
\begin{eqnarray}
\Gamma = b_0 \left ( \frac{1}{M}
 \left(
	j + \begin{matrix}\frac{1}{2}\end{matrix} -  \sqrt{ (j+ \begin{matrix}\frac{1}{2}\end{matrix})^2 - M^2}
\right) 
	+ \frac{j+1 - \sqrt{(j+\begin{matrix}\frac{1}{2}\end{matrix})^2 - M^2 } } {2 (j+1)} a_0 z + O(z^2) 
\right).
\label{section4 equation 40}
\end{eqnarray}
For $M = 0$,
\begin{eqnarray}
\triangle = b_0 \left ( 
1 + \frac{a_0^2}{16(j+1)} z^2 + O(z^3) 
\right),
\label{section4 equation 41}
\end{eqnarray}
\begin{eqnarray}
\Gamma = b_0 \left ( 
\frac{a_0}{4(j + 1) } z + \frac{a_1}{2(2j+3)} z^2 + 0(z^3)
\right).
\label{section4 equation 42}
\end{eqnarray}

Having established that there is a solution of (\ref{section4 equation 37}) and (\ref{section4 equation 38})
that is finite at $r = 0$ we now show that this solution is not
square-integrable. 
There is also a solution that is too singular at the origin to satisfy 
(\ref{section4 equation 25}); we therefore ignore it here.  Let
\begin{eqnarray}
\triangle & = & z^{-j -\frac{1}{2}} \delta, \nonumber\\
\Gamma & = & z^{-j -\frac{1}{2}} \gamma, \nonumber\\
\lambda &=& \sqrt{ (j +\begin{matrix}\frac{1}{2}\end{matrix})^2 - M^2 }. 
\label{section4 equation 43}
\end{eqnarray}
Then (\ref{section4 equation 37}), (\ref{section4 equation 38}) reduce to
\begin{eqnarray}
z \frac{d\delta}{dz} - \lambda \delta = (M + \begin{matrix}\frac{1}{2}\end{matrix} a z) \gamma,
\label{section4 equation 44}
\end{eqnarray}
\begin{eqnarray}
z \frac{d\gamma}{dz}+ \lambda \gamma= (M + \begin{matrix}\frac{1}{2}\end{matrix} a z) \delta.
\label{section4 equation 45}
\end{eqnarray}
Multiply (\ref{section4 equation 44}) by  $\delta$ , (\ref{section4 equation 45}) by $\gamma$  and subtract:
\begin{eqnarray}
\frac{d}{dr} (\delta^2 - \gamma^2)  = \frac{4\lambda}{r} (\gamma^2 + \delta^2),
\label{section4 equation 46}
\end{eqnarray}
where
\begin{eqnarray}
\delta^2 - \gamma^2 = 4 \left [ \lambda (f^2 - r^4 g^2) - 2M r^2 f g \right ] r^{4j+2},
\label{section4 equation 47}
\end{eqnarray}
and
\begin{eqnarray}
\delta^2 + \gamma^2 = 2(2j + 1) \left [ f^2 + (r^2 g)^2 \right ] r^{4j+2}.
\label{section4 equation 48}
\end{eqnarray}
From (\ref{section4 equation 25}), if   $\psi^- \in L^2$  then $f$, $r^2 g \underset{{r \rightarrow \infty}}{\sim}    r^{-2j-1 - \epsilon}$ , $\epsilon   >0$, in which case
$\underset{{r = \infty}}{\lim}(\delta^2  -\gamma^2  ) = 0$. Because  $\triangle$ , $\Gamma$   are finite at $r = 0$, $\gamma$  ,$\delta   = O(r^{2j+1})$   as
$r\rightarrow 0$. Hence, integration of (\ref{section4 equation 46}) using (\ref{section4 equation 48}) gives
\begin{eqnarray}
\sqrt{ (j + \begin{matrix}\frac{1}{2}\end{matrix})^2 - M^2 } \int_0^\infty dr r^{4j + 1} \left [ f^2 + (r^2g)^2 \right] = 0.
\label{section4 equation 49}
\end{eqnarray}
But this is impossible for $|M|< j + \begin{matrix}\frac{1}{2}\end{matrix}$. Therefore, the assumption that
 $\psi^- \in L^2$  is false.

    Summarizing, it has been shown that all $L^2$  zero modes of $\displaystyle{\not}D$ have
positive chirality and that these only occur when $M = -j-\begin{matrix}\frac{1}{2}\end{matrix}$  and for
values of $j$ satisfying $\nu  > 2j + 2$.

    These results raise an interesting problem. The main result
of \cite{bib_15} 
is
\begin{eqnarray}
\frac{\nu^2}{2} = n_+ - n_- + \frac{1}{\pi} \sum_l \mu (l) \left [ \delta_l^+ (0) - \delta_l^-(0) \right ] ,  
\label{section4 equation 50}
\end{eqnarray}
where $n_\pm$  are the number of positive and negative chirality $L^2$  zero
modes, $\delta_l^\pm   (0)$ are the zero-energy scattering phase shifts for
$H_{\pm}$  in (\ref{section2 equation 8}), $ \mu (l)$ is a weight factor, and $l$ are the quantum numbers required to specify the phase shifts discussed in Sec. \hspace{-2mm}VI. We have just shown
that $n_- = 0$. Suppose $\nu  = 3$. Due to the condition for a $L^2$  zero mode
derived above only $j = 0$, $M = -\frac{1}{2}$  , $m = 0$ are allowed. So $n_+  = 1$, and
it must follow that
\begin{equation}
\frac{9}{2} = 1 + \frac{1}{\pi} \sum_l \mu (l) \left[ \delta_l^+ (0) - \delta_l^- (0) \right ].
\label{section4 equation 51}
\end{equation}
Verification of this and (\ref{section4 equation 50}) here would take us too far afield.


\section{Scattering states}
    Having established that there are no negative chirality
zero modes it cannot be concluded that the $m = 0$ limit of
the last term in (\ref{section2 equation 13}) is zero.  Equation (\ref{newsection4 equation 1})
when combined with (\ref{section3 equation 19}) and (\ref{section4 equation 50}) caution against this. They
demonstrate that a particular zero-mass limit receives 
contributions from the scattering states of $H_-$   in (\ref{section2 equation 8}). There seems
to be no alternative to actually calculating the low-energy
scattering states of $H_-$  before deciding whether the $m = 0$ limit of
the last term in (\ref{section2 equation 13}) is zero.

    Because $\displaystyle{\not}D$ is anti-Hermitian we look for eigenstates of the form
\begin{equation}
\displaystyle{\not}D \psi = ik \psi.
\label{section5 equation 1}
\end{equation}
Decomposing  $\psi$  into its positive and negative chirality components
and using (\ref{section4 equation 1}) gives
\begin{equation}
D D^\dagger \psi^+ = k^2 \psi^+,
\label{section5 equation 2}
\end{equation}
\begin{equation}
D^\dagger D \psi^- = k^2 \psi^-.
\label{section5 equation 3}
\end{equation}
To get the scattering states $\psi^-$  it is easier to calculate $\psi^+$   and then
use $D^\dagger \psi^+  = -ik \psi^-$. In the representation (\ref{section2  equation 7}) the Zeeman term, $\frac{1}{2}\sigma F$    ,
is diagonal in the positive chirality sector, and so $D D^\dagger = H_+$  has
the form
\begin{equation}
D D^\dagger = \left ( 
\begin{array}{cc}
H_{\frac{1}{2}} & 0 \\
0 & H_{-\frac{1}{2}} 
\end{array}
\right ),
\label{section5 equation 4}
\end{equation}
where the subscripts on    $H$ denote the eigenvalues of $S_3^{(1)}$   in (\ref{section3 equation 6}).
In (\ref{section4 equation 3}) let
\begin{equation}
\sqrt{\frac{2j+1}{2\pi^2} } r^{-2j - \frac{3}{2} } \rho_{\pm \frac{1}{2}} = (j \mp M + \begin{matrix}\frac{1}{2}\end{matrix})^{\frac{1}{2}} F \mp (j \pm M +\begin{matrix}\frac{1}{2}\end{matrix})^{\frac{1}{2}} G, 
\label{section5 equation 5}
\end{equation}
and decompose $\psi^+$   into its upper and lower components:
\begin{eqnarray}
\psi_{\frac{1}{2}}^+ &= & \sqrt{ \frac{2j+1}{2\pi^2}} \left ( 
\begin{array}{c}
D_{M-\frac{1}{2}, m}^j (\hat{x}) \\
0
\end{array}
\right ) \frac{\rho_{\frac{1}{2}}(r)}{r^{\frac{3}{2}}},\nonumber\\
\psi_{- \frac{1}{2}}^+ &= & \sqrt{ \frac{2j+1}{2\pi^2}} \left ( 
\begin{array}{c}
0 \\
D_{M+\frac{1}{2}, m}^j (\hat{x}), 
\end{array}
\right ) \frac{\rho_{- \frac{1}{2}} (r)}{r^{\frac{3}{2}}},
\label{section5 equation 6}
\end{eqnarray}
where $\hat{x}\cdot \hat{ x} = 1$. Substituting Eqs.(\ref{section5 equation 6}) in turn in (\ref{section5 equation 2}) gives
\begin{eqnarray}
\left [ - \frac{d^2}{dr^2} + \frac{ (2j+1)^2 - \begin{matrix}\frac{1}{4}\end{matrix}}{r^2} + (4M \pm 2) a + r^2 a^2 \pm r \frac{da}{dr} \right ] \rho_{\pm \frac{1}{2}} = k^2 \rho_{\pm \frac{1}{2}} .  
\label{section5 equation 7}
\end{eqnarray}
Equation (\ref{section5 equation 7}) has to be supplemented by appropriate boundary
conditions. For $r > R$ according to (\ref{section3 equation 3}) $a =  \nu/r^2$ . Let
\begin{eqnarray}
\rho_{\pm \frac{1}{2} } = r^{\frac{1}{2}} f_\pm.
\label{section5 equation 8}
\end{eqnarray}
Then (\ref{section5 equation 7}) becomes for $r > R$
\begin{eqnarray}
f_\pm^{''} + \frac{1}{r} f_{\pm}^{'} + \left (k^2 - \frac{(2j+1)^2 + 4M \nu + \nu^2}{r^2} \right ) f_\pm = 0,
\label{section5 equation 9}
\end{eqnarray}
whose general solution is a superposition of Hankel functions
\begin{eqnarray}
f_{\pm} = \alpha_\pm H_\lambda^{(1)} (kr) + \beta_\pm H_\lambda^{(2)} (kr), 
\label{section5 equation 10}
\end{eqnarray}
with
\begin{eqnarray}
\lambda = \left [ 
(2j+1)^2  + 4M\nu + \nu^2 
\right ]^{\frac{1}{2}}.
\label{section5 equation 11}
\end{eqnarray}
Choosing $\alpha_\pm$   ,$\beta_\pm$    so that
\begin{eqnarray}
\rho_{EjM,\pm \frac{1}{2}} (r) \underset{kr >> 1}{\sim} \sqrt{ \frac{1}{\pi k}} \cos \left ( kr - \begin{matrix}\frac{\pi}{2}\end{matrix}  (2j + 1) + \delta_{jM,\pm\frac{1}{2}}^+ (k) - \begin{matrix}\frac{\pi}{4}\end{matrix}\right ),
\label{section5 equation 12}
\end{eqnarray}
gives for $r > R$
\begin{eqnarray}
\rho_{E\alpha} (r) &=& \sqrt{ \frac{r}{8}} \left ( e^{i\left( \frac{\pi \lambda}{2} - \frac{\pi}{2}(2j + 1) + \delta_{\alpha}^+(k) \right)}H_\lambda^{(1)} (kr) \right. \nonumber\\
&& \left. + e^{-i\left( \frac{\pi \lambda}{2} - \frac{\pi}{2}(2j + 1) + \delta_{\alpha}^+(k) \right)}H_\lambda^{(2)} (kr) \right ),
\label{section5 equation 13}
\end{eqnarray}
where $E = k^2$  and $\alpha$  denotes $j$, $M$, $\pm \frac{1}{2}$. The superscript on  $\delta_\alpha^+$  is a
reminder that these are positive chirality phase shifts. The solutions
(\ref{section5 equation 13}) are to be joined to the solutions of (\ref{section5 equation 7}) for $r < R$. This
will determine the phase shifts. Equation (\ref{section5 equation 13}) fixes the
normalization so that
\begin{equation}
\int_0^\infty dr \rho_{E\alpha} (r) \rho_{E'\alpha} (r) = \delta (E - E').
\label{section5 equation 14}
\end{equation}
Then  $\psi_{\pm \frac{1}{2}}^+$  in (\ref{section5 equation 6}) have the overall normalization
\begin{equation}
\left( \psi_{E\beta}^+, \psi_{E'\beta'}^+ \right )  = \delta_{\beta \beta'} \delta(E - E'),
\label{section5 equation 15}
\end{equation}
where $\beta$   represents $j$, $M$ , $m$, $\pm \frac{1}{2}$.

    The calculation of the low-energy phase shifts is outlined in
Appendix B. Define the energy-dependent part of $\delta_\alpha^+$   by
\begin{equation}
\triangle_\alpha^+ (k) = \frac{\pi \lambda}{2} - \frac{\pi}{2} (2j + 1) + \delta_\alpha^+ (k), \mod \pi,
\label{section5 equation 16}
\end{equation}
and denote the expansion in powers of $k$ of the logarithmic derivative
of the interior radial wave function at $r = R$ by
\begin{equation}
\left ( \frac{r \partial_r \rho_{E\alpha}}{\rho_{E\alpha} } \right )_R = 
\gamma_\alpha - (kR)^2 \Gamma_\alpha + O(kR)^4.
\label{section5 equation 17}
\end{equation}
The coefficients $\gamma_\alpha$  , $\Gamma_\alpha$  are defined in Appendix B. Then for $|M| \not= j+\frac{1}{2}$
\begin{equation}
\tan \triangle_\alpha^+ = - \frac{\pi}{\lambda \Gamma^2 (\lambda) } \frac{\gamma_\alpha - \lambda - \begin{matrix}\frac{1}{2}\end{matrix}}{\gamma_\alpha + \lambda -\begin{matrix}\frac{1}{2}\end{matrix}} 
\left ( \frac{kR}{2} \right )^{2\lambda} 
\left ( 1 + O \left [  (kR)^2, (kR)^{2\lambda} \right ] \right ),
\label{section5 equation 18}
\end{equation}
with  $\lambda > 1$ for $\nu  > 0$. There are several special cases to consider.

    Case 1: $M = j+\frac{1}{2}$   and hence $\lambda  = 2j+1+\nu$  . From (\ref{section5 equation 6}) the only
phase shift in this case is $\delta_{jj+\frac{1}{2}, \frac{1}{2}}^+ (k)$       and
\begin{equation}
\tan \triangle_{jj+\frac{1}{2}, \frac{1}{2}}^+ = - \frac{\pi}{\Gamma^2 ( 1 + \lambda )} \left ( 
\frac{1}{\lambda + 1} - 2 \Gamma_{jj+\frac{1}{2}, \frac{1}{2}}
\right ) \left (
\frac{kR}{2}
\right )^{2 \lambda + 2} \left ( 1 + O\left[ (kR)^2, (kR)^{2\lambda - 2} \right ]\right ). 
\label{section5 equation 19}
\end{equation}

    Case 2: $M = -j-\frac{1}{2}$  and hence $\lambda  = |2j+1-\nu  |$. From (\ref{section5 equation 6}) the only
phase shift in this case is $\delta_{j, -j-\frac{1}{2}, -\frac{1}{2}}^+$.\\
Case 2.1: $2j< \nu <2j+1$ with $0<\lambda  <1$,
\begin{equation}
\tan \triangle_{j, -j-\frac{1}{2}, -\frac{1}{2}}^+ = - \frac{\pi}{\lambda \Gamma^2 ( \lambda )} \left ( 
\frac{\frac{1}{\lambda + 1} - 2 \Gamma_{j, -j-\frac{1}{2}, -\frac{1}{2}}} { 2 j + 1 - \nu }
\right ) \left (
\frac{kR}{2}
\right )^{2 \lambda + 2} \left ( 1 + O\left[ (kR)^2, (kR)^{2\lambda} \right ]\right ). 
\label{section5 equation 20}
\end{equation}
Case 2.2: $2j+1<\nu <2j+2$ with $0< \lambda <1,$
\begin{equation}
\tan \triangle_{j, -j-\frac{1}{2}, -\frac{1}{2}}^+ = \tan \pi \lambda + O\left[(kR)^{2 - 2\lambda}\right]. 
\label{section5 equation 21}
\end{equation}
When $\lambda = \frac{1}{2} + \epsilon, |\epsilon| << 1$, (\ref{section5 equation 21}) becomes
\begin{eqnarray}
\tan \triangle_{j, -j-\frac{1}{2},-\frac{1}{2}}^+ = - \frac{
1 + 2 \epsilon + O(kR)^2 
} { 
\pi \epsilon + 2 (\Gamma_{j, -j-\frac{1}{2},-\frac{1}{2}} - 1) 
\left ( \frac{kR}{2} \right )^{1 - 2 \epsilon} + O (\epsilon^2, \epsilon (kR)^{1 - 2\epsilon}, (kR)^{3 - 2 \epsilon} ) }. 
\end{eqnarray}
Case 2.3: $0< \nu <2j$ with  $\lambda > 1$,
\begin{equation}
\tan \triangle_{j, -j-\frac{1}{2}, -\frac{1}{2}}^+ = - \frac{\pi}{\lambda \Gamma^2 ( \lambda )} \left ( 
\frac{\frac{1}{\lambda + 1} - 2 \Gamma_{j, -j-\frac{1}{2}, -\frac{1}{2}}} { 2 j + 1 - \nu }
\right ) \left (
\frac{kR}{2}
\right )^{2 \lambda + 2} \left ( 1 + O(kR)^2\right ). 
\label{section5 equation 22}
\end{equation}
Case 2.4: $\nu >2j+2$ with $\lambda  >1$,
\begin{equation}
\tan \triangle_{j, -j-\frac{1}{2}, -\frac{1}{2}}^+ =  \frac{\pi}{\lambda \Gamma^2 ( \lambda )} \left ( 
\frac{ 2 j + 1 - \nu }{2 \Gamma_{j, -j-\frac{1}{2}, -\frac{1}{2}} +  \frac{1}{\lambda - 1}   } 
\right ) \left (
\frac{kR}{2}
\right )^{2 \lambda - 2} \left ( 1 + O\left[ (kR)^2 , (kR)^{2\lambda - 2} \right ] \right ). 
\label{section5 equation 23}
\end{equation}
Case 2.5: $\nu   = 2j+1$ with $\lambda  = 0$,
\begin{equation}
\tan \triangle_{j, -j-\frac{1}{2}, -\frac{1}{2}}^+ = -\pi\left( 
1-2\Gamma_{j,-j-\frac12,-\frac12}\right)\left(\frac{kR}{2}\right)^2
\left(1+O\left[(kR)^2\ln(kR)\right]\right). 
\label{section5 equation 24}
\end{equation}
Case 2.6: $ \nu = 2j+2$ with $\lambda  = 1$,
\begin{equation}
\tan \triangle_{j, -j-\frac{1}{2}, -\frac{1}{2}}^+ = \frac{ 
\frac{\pi}{2}\left(1+O\left[(kR)^2\right]\right) 
}{
\ln \left(\frac{kR}{2}\right) + \gamma_E - \Gamma_{j, -j-\frac{1}{2}, -\frac{1}{2}} + O\left[ (kR)^2 \ln(kR)\right]
},
\label{section5 equation 25}
\end{equation}
where $\gamma_E$ is Euler's constant $0.577\dots$ \\
Case 2.7: $\nu  = 2j$ with  $\lambda = 1$,
\begin{equation}
\tan \triangle_{j, -j-\frac{1}{2}, -\frac{1}{2}}^+ = - \pi \left ( 
\begin{matrix} \frac{1}{2}\end{matrix} - 2 \Gamma_{j, -j-\frac{1}{2}, -\frac{1}{2}} 
\right ) \left (
\frac{kR}{2}
\right )^{4} \left(1+O\left[(kR)^2\right]\right). 
\label{section5 equation 24}
\end{equation}

    Although it is not required for the analysis here there is a
compact relation between the interior wave functions and the phase
shifts that ought to be mentioned, namely for $|M| \not= j + \frac{1}{2}$ 
\begin{equation}
2 \pi \int_0^R \frac{ds}{s} \frac{d}{ds} (s^2 a) \rho_{jM\frac{1}{2}} (s) \rho_{jM,-\frac{1}{2}} (s) = \sin \left ( 
\delta_{jM\frac{1}{2}}^+ (k) - \delta_{jM, -\frac{1}{2}}^+ (k) \right). 
\label{section5 equation 27}
\end{equation}
This is easily obtained by going back to (\ref{section5 equation 7}) and noting that
\begin{equation}
\left ( \rho_{\frac{1}{2}}^{'} \rho_{-\frac{1}{2}} - \rho_{-\frac{1}{2}}^{'} \rho_{\frac{1}{2} }\right )' = \frac{2}{r} (r^2 a)' \rho_{\frac{1}{2}} \rho_{-\frac{1}{2}} .
\label{section5 equation 28}
\end{equation}
For $r > R$ the right-hand side of (\ref{section5 equation 28}) vanishes. The constant
$\rho_{\frac{1}{2}}^{'}  \rho_{-\frac{1}{2}} - \rho_{-\frac{1}{2}}^{'}  \rho_{\frac{1}{2}}$   in the region $r > R$ can be calculated using (\ref{section5 equation 12}). Then
integrating (\ref{section5 equation 28}) from $0$ to $R $ gives (\ref{section5 equation 27}). It holds for all energies.

    We now proceed to get the negative chirality scattering
states, in particular $f $ and $g$ in (\ref{section3 equation 16}) by calculating
$D^{\dagger} \psi_{\pm \frac{1}{2}}^{+}   = -ik \psi_{\pm\frac{1}{2}}^{-}$   . This results in two orthogonal states
since
\begin{eqnarray}
\left( \psi_{\frac{1}{2}}^{-}, \psi_{-\frac{1}{2}}^{-} \right ) &=& \frac{1}{k^2} \left ( \psi_{\frac{1}{2}}^{-}, D^{\dagger} D \psi_{-\frac{1}{2}}^{-} \right) 
= \frac{1}{k^2} \left ( D \psi_{\frac{1}{2}}^{-} , D \psi_{-\frac{1}{2}}^{-} \right) 
= \left ( \psi_{\frac{1}{2}}^{+}, \psi_{-\frac{1}{2}}^{+} \right) = 0.
\label{section5 equation 29}
\end{eqnarray}
The result is
\begin{eqnarray}
\psi_{EjMm\frac{1}{2}}^{-} (x) =\frac{1}{\sqrt{2\pi^2} kr^{\frac{3}{2}} } \sqrt {\frac{j-M+\frac{1}{2} }{2j+1} }
\left (
\begin{array}{c}
(j+m)^{\frac{1}{2}} D_{M,m-\frac{1}{2}}^{j - \frac{1}{2}} (\hat{x}) \\
(j-m)^{\frac{1}{2}} D_{M,m+\frac{1}{2}}^{j - \frac{1}{2}} (\hat{x})
\end{array}
\right )
\left ( \frac{d}{dr} - ar + \frac{2j + \begin{matrix}\frac{1}{2}\end{matrix}}{r} \right ) \rho _{EjM\frac{1}{2}} (r) \nonumber\\
 - 
 \frac{1}{\sqrt{2\pi^2} kr^{\frac{3}{2}} } \sqrt {\frac{j+M+\frac{1}{2} }{2j+1} }
\left (
\begin{array}{c}
-(j-m+1)^{\frac{1}{2}} D_{M,m-\frac{1}{2}}^{j + \frac{1}{2}} (\hat{x}) \\
(j+m+1 )^{\frac{1}{2}} D_{M,m+\frac{1}{2}}^{j + \frac{1}{2}} (\hat{x})
\end{array}
\right )
\left ( \frac{d}{dr} - ar - \frac{2j + \frac{3}{2}}{r} \right ) \rho _{EjM\frac{1}{2}} (r),\nonumber\\
\label{section5 equation 30}
\end{eqnarray}
\begin{eqnarray}
\psi_{EjMm,-\frac{1}{2}}^{-} (x) = \frac{1}{\sqrt{2\pi^2} kr^{\frac{3}{2}} } \sqrt {\frac{j+M+\frac{1}{2} }{2j+1} }
\left (
\begin{array}{c}
(j+m)^{\frac{1}{2}} D_{M,m-\frac{1}{2}}^{j - \frac{1}{2}} (\hat{x}) \\
(j-m)^{\frac{1}{2}} D_{M,m+\frac{1}{2}}^{j - \frac{1}{2}} (\hat{x})
\end{array}
\right )
\left ( \frac{d}{dr} + ar + \frac{2j + \begin{matrix}\frac{1}{2}\end{matrix}}{r} \right ) \rho _{EjM,-\frac{1}{2}} (r) \nonumber\\
 + 
 \frac{1}{\sqrt{2\pi^2} kr^{\frac{3}{2}} } \sqrt {\frac{j-M+\frac{1}{2} }{2j+1} }
\left (
\begin{array}{c}
-(j-m+1 )^{\frac{1}{2}} D_{M,m-\frac{1}{2}}^{j + \frac{1}{2}} (\hat{x}) \\
(j+m+1)^{\frac{1}{2}} D_{M,m+\frac{1}{2}}^{j + \frac{1}{2}} (\hat{x})
\end{array}
\right )
\left ( \frac{d}{dr} + ar - \frac{2j + \frac{3}{2}}{r} \right ) \rho _{EjM, -\begin{matrix}\frac{1}{2}\end{matrix}} (r). \nonumber\\
\label{section5 equation 31}
\end{eqnarray}
These states are normalized so that
\begin{equation}
\left ( \psi_{E\beta}^{-}, \psi_{E'\beta'}^{-} \right )  = \delta_{\beta \beta'} \delta{(E-E')},
\label{section5 equation 32}
\end{equation} 
where $\beta$  represents $j$, $M$, $m$, $\pm\frac{1}{2}$. Because there are no $L^2$  zero modes in
the negative chirality sector we expect that the scattering states (\ref{section5 equation 30}) and
(\ref{section5 equation 31}) form a complete set:
\begin{eqnarray}
\sum_{j=0}^\infty \sum_{M=-j-\begin{matrix}\frac{1}{2}\end{matrix}}^{j+\frac{1}{2}} \sum_{m = -j}^j \int_0^\infty dE 
\left [ 
\psi_{EjMm\frac{1}{2}}^{-} (x) \psi_{EjMm\frac{1}{2}}^{-\dagger} (x') + 
\psi_{EjMm-\frac{1}{2}}^- (x) \psi_{EjMm,-\frac{1}{2}}^{-\dagger} (x') \right ] = \delta (x - x'){1\hskip-0,27em{\rm I}}_2.
\nonumber\\
\label{section5 equation 33}
\end{eqnarray}

\section{$\Delta_-$ at Low Energy }
The exact negative chirality propagator is 
\begin{eqnarray}
\triangle_-(x,x') = \sum_\alpha \int_0^\infty dk^2 \frac{\psi_{E\alpha}^- (x) \psi_{E\alpha}^{-\dagger} (x')}{k^2 + m^2},
\label{section6 equation 2}
\end{eqnarray}
with $\psi_{E\alpha}^{-}$     given by (\ref{section5 equation 30}), (\ref{section5 equation 31}) and $\alpha  = jMm, \pm \frac{1}{2}$  . 
Now suppose $\triangle_-  (x,x')$ is divided into its low and high energy
 parts by replacing the integral in (\ref{section6 equation 2}) by $\int_0^{\Lambda^2}   + \int_{\Lambda^2}^\infty$, with $\Lambda  R<<1$.
 Then our objective is to show that the low energy propagator
 has only minor deviations from the free propagator. This turns out to
be the case except when $\nu = 2j+2$ which results in a benign logarithmic
 mass singularity. The high energy propagator poses no obstacle to
 the $m=0$ limit in (\ref{section2 equation 13}) and is well-defined due to the assumed regularity
 of $A_\mu$   at the origin.

     In order to proceed we replace the differential equation (\ref{section5 equation 7})
with the integral equation
\begin{eqnarray}
\rho_{\pm} (r) &=& A_{\pm} \sqrt{\frac{r}{2}} J_{2j+1} (kr) + \frac{\pi}{2} \sqrt{r} \int_0^r dr' \sqrt{r'}\nonumber\\
&& \times \left[
J_{2j+1} (kr')Y_{2j+1} (kr) - J_{2j+1} (kr) Y_{2j + 1} (kr')
\right] V_\pm (r') \rho_\pm (r'), 
\label{section6 equation 4}
\end{eqnarray}
where $\rho_\pm$  represents  $\rho_{EjM,\pm\frac{1}{2}}      , A_\pm$  are constants to be determined and
\begin{eqnarray}
V_\pm = (4M \pm 2) a + r^2 a^2 \pm r \frac{da}{dr} .
\label{section6 equation 5}
\end{eqnarray}
By differentiating (\ref{section6 equation 4}) it can be verified that (\ref{section5 equation 7}) results. To
 fix $A_\pm$   require that $\rho_\pm$   join smoothly to the outgoing wave solution
 (\ref{section5 equation 13}) at $r=R$ with $\delta_\alpha^+$    replaced by its energy dependent part
 defined in (\ref{section5 equation 16}). Then
\begin{eqnarray}
\rho_\pm (R) = \sqrt{\frac{R}{2}} \left( J_\lambda (kR) \cos \triangle_\alpha^+ (k) - Y_\lambda (kR) \sin \triangle_\alpha^+ (k) \right), 
\label{section6 equation 6}
\end{eqnarray}
together with (\ref{section6 equation 4}) at $r=R$ determine $A_\pm$.

     An upper bound on $ \rho_\pm (r)$ for $0\leq r \leq R$ will now be obtained. Starting
with (2.60), (12.134) and (12.136a) in \cite{bib_20} 
deduce that for $z>0$ and
for fixed values of $j$
\begin{eqnarray}
|J_{2j+1} (z) |  \leq  \frac{C_J \left ( \frac{z}{2} \right )^{2j+1} }{ (2j+1)! (1 + z) ^{2j+\frac{3}{2}} },
\label{section6 equation 7}
\end{eqnarray}

\begin{eqnarray}
|H_{2j+1}^{(1)} (z) |  \leq  C_H \sqrt{\frac{2}{\pi z} } \left ( \frac{1+z}{z} \right )^{2j+\frac{1}{2}},
\label{section6 equation 8}
\end{eqnarray}
where the constants $C_J$  , $C_H$  depend on $j$. From these results it follows
that for $z \geq z' >0$
\begin{eqnarray}
\sqrt{z z'} \Big| J_{2j+1}(z') Y_{2j+1} (z) - J_{2j+1} (z) Y_{2j+1} (z') \Big| \leq C z \left ( \frac{z}{z'} \right )^{2j+1}, 
\label{section6 equation 9}
\end{eqnarray}
with $C$ of order one. Now iterate (\ref{section6 equation 4}) and let
\begin{eqnarray}
\rho_\pm (r) &=& \sum_0 \rho_\pm^{(n)} (r), \nonumber \\
\rho_\pm^{(0)} (r) & = & A_\pm \sqrt{\frac{r}{2} } J_{2j+1}(kr),\nonumber \\
| \rho_\pm^{(n)} (r) | & = & r \psi_\pm ^{(n)} (r). 
\label{section6 equation 10}
\end{eqnarray}
From (\ref{section6 equation 4}) and (\ref{section6 equation 9}) the $n^{th}$ iterate satisfies
\begin{eqnarray}
\psi_\pm^{(n)} (r) \leq \frac{\pi C}{2} \int_0^r dr' |V_\pm (r')| \left (\frac{r}{r'} \right )^{2j+1} r' \psi_\pm^{(n-1)} (r').
\label{section6 equation 11}
\end{eqnarray}
Since $0<r<R$ and we assume $kR<<1$, then (\ref{section6 equation 10}) gives
\begin{eqnarray}
|\rho_\pm^{(0)} (r)| = r \psi_\pm^{(0)} (r) \leq \sqrt{\frac{r}{2}} \left ( \frac{kr}{2} \right )^{2j+1} \frac{|A_\pm|}{(2j+1)!}.
\label{section6 equation 12}
\end{eqnarray}
Thus (\ref{section6 equation 11}) and (\ref{section6 equation 12}) give
\begin{eqnarray}
\psi_\pm^{(n)} &\leq& \frac{\left ( \frac{\pi C}{2}\right )^n }{\sqrt{2r}} \left ( \frac{kr}{2} \right)^{2j+1} \frac{|A_\pm |}{(2j+1)!} r^n \int_0^r dr_n \dots \int_0^{r_2} dr_1 |V_\pm (r_1) \dots V_\pm (r_n) | \nonumber \\
&= & \frac{|A_\pm |}{\sqrt{2r} (2j+1)!}\left ( \frac{kr}{2} \right)^{2j+1} \left[ \frac{\pi C r}{2} \int_0^r ds | V_\pm (s) | \right ] ^n / n!. 
\label{section6 equation 13}
\end{eqnarray}
By (\ref{section6 equation 10})
\begin{eqnarray}
|\rho_\pm (r)| \leq \sqrt{\frac{r}{2}} \frac{|A_\pm|}{(2j+1)!} \left(
	\frac{kr}{2} 
\right)^{2j+1} \exp \left( 
	\frac{\pi Cr}{2} \int_0^r ds | V_\pm (s) | 
\right), 
\label{section6 equation 14}
\end{eqnarray}
valid for $0\leq r \leq R$, $kR<<1$.

It remains to estimate the constants $A_\pm$. Suppose $|M|\not= j+\frac{1}{2}$.
From (\ref{section6 equation 4}), (\ref{section6 equation 6}) and (\ref{section5 equation 18}) with $kR<<1$ obtain
\begin{eqnarray}
A_\pm &=& \left ( \frac{kR}{2} \right)^{\lambda -2j -1} \left ( 1 - \frac{\Gamma (2j+2)}{\Gamma(\lambda +1)} \frac{\gamma_\pm - \lambda - \begin{matrix}\frac{1}{2}\end{matrix}}{ \gamma_\pm + \lambda - \begin{matrix}\frac{1}{2}\end{matrix}} + O(kR)^2 \right ) \nonumber \\
&& - \frac{\Gamma (2j+1)}{\sqrt{2}} \int_0^R dr \sqrt {r} \left (1 - \left (\frac{r}{R} \right )^{4j+2} + O(kR^2) \right) V_\pm (r) (\begin{matrix}\frac{1}{2}\end{matrix} kr)^{-2j-1} \rho_\pm (r).
\label{section6 equation 15}
\end{eqnarray}
Because $k^2$  is an analytic perturbation of  $\rho_\pm$ in (\ref{section5 equation 7}) make the
expansion
\begin{eqnarray}
\rho_\pm (r) = \frac{\left ( \frac{kR}{2} \right)^\lambda }{ (2j+1)!} \left ( \rho_{0\pm} (r) + \rho_{2\pm} (r) k^2 + O (k^4) \right ), 
\label{section6 equation 16}
\end{eqnarray}
for $0 \leq r \leq R$. Then to leading order in $k$
\begin{eqnarray}
A_\pm &=& \left ( \frac{kR}{2} \right )^{\lambda -2 j - 1 } 
\left [ 
1 - \frac{\Gamma (2j+2)}{\Gamma (\lambda +1)} \frac{(\gamma_\pm - \lambda -\begin{matrix}\frac{1}{2}\end{matrix}) } {(\gamma_\pm + \lambda -\begin{matrix}\frac{1}{2}\end{matrix}) }  \right. \nonumber \\
&& \left. - \frac{1}{\sqrt{2}} \frac{1}{2j+1} \int_0^R dr \sqrt{r} \left ( 1 - \left (\frac{r}{R} \right )^{4j+2} \right ) V_\pm (r) \left ( \frac{R}{r} \right )^{2j+1} 
\rho _{0 \pm} (r) \right ].  
\label{section6 equation 17}
\end{eqnarray}
Substitution of (\ref{section6 equation 17}) and (\ref{section6 equation 16}) into (\ref{section6 equation 4}) gives an integral
equation for  $\rho_{0\pm}  (r)$ whose solution we do not require here. The main
conclusion is that for $kR<<1$, $|M|\not=j+\frac{1}{2}$
\begin{eqnarray}
A_\pm = \left ( \frac{kR}{2} \right )^{\lambda - 2j - 1} N_\pm (j, M, \nu),
\label{section6 equation 18}
\end{eqnarray}
where $N_\pm (j, M, \nu)$ is $k-$independent. The centrifugal barrier term in (\ref{section5 equation 7})
for large $j$ (certainly $j>> \nu$)  will cause $\rho_\pm$   to approach the noninteracting solution
$\sqrt{\frac{r}{2}}   J_{2j+1}   (rk)$. Hence, for large $j$, $ \lambda  \rightarrow  2j+1, \gamma_\pm  \rightarrow  2j+  \frac{3}{2}, N_\pm  (j, M, \nu ) \rightarrow 1$
and so $A_\pm \rightarrow  1$. Equations (\ref{section6 equation 14}) and (\ref{section6 equation 18}) give the explicit upper
bound
\begin{eqnarray}
| \rho_\pm (r) | \leq \frac{1}{(2j+1)!} \sqrt{\frac{r}{2}} \left ( \frac{r}{R} \right )^{2j+1} \left ( \frac{kR}{2} \right )^\lambda | N_\pm (j, M, \nu) | e^{\frac{1}{2} \pi C r \int_0^R ds | V_\pm (s)| } , 
\label{section6 equation 19}
\end{eqnarray}
for $0\leq r \leq R$, $kR<<1$, $|M|\not=j+\frac{1}{2}$. The $k-$dependence of this bound is
consistent with (\ref{section5 equation 27}).

When $M=j+\frac{1}{2}$, case 1, Sec. VI, only $\rho_+$   is relevant and there is
no change in its overall $k$ dependence. When $M=-j-\frac{1}{2}$   only $\rho_-$   is
relevant, and the largest modification of (\ref{section6 equation 19}) occurs in case 2.2
when $2j+1< \nu <2j+2$ with 0$< \lambda <1$. Repeating the above analysis gives the
same result as (\ref{section6 equation 19}) except that the factor $(kR/2)^\lambda$  is replaced
with $(kR/2)^{-\lambda}$   and $N_-$  is replaced with a new constant $\tilde{N}_-$. Thus, for
case 2.2 (\ref{section6 equation 16}) is replaced with
\begin{eqnarray}
\rho_- (r) = \frac{\left( \frac{kR}{2} \right)^{-\lambda} }{(2j+1)!} \left (\rho_{0-} + \rho_{2-} (r)k^2 + O(k^4) \right ). 
\label{section6 equation 20}
\end{eqnarray}
The remaining cases when $M=-j-\frac{1}{2}$   result in less singular $k-$factors
than $(kR)^{-\lambda}$.

Now it is evident that the overall $k-$dependence of $\rho_\pm  (r)$ is
not changed by differentiating it with respect to $r$. Therefore, the
leading small $k-$dependence of the radial wave functions in (\ref{section5 equation 30})
and (\ref{section5 equation 31}) remains $(kR)^\lambda$, $\lambda  >1$ for $M\not= -j-\frac{1}{2}$. Because of the factor
$k^{-1}$   multiplying them the negative chirality wave functions $\psi_{EjM,\pm\frac{1}{2}}^-$
fall off as $(kR)^{\lambda -1}$  as $k\rightarrow 0$ for $M \not= -j- \frac{1}{2}$, $0\leq r \leq R$. This statement was
verified by deriving integral equations for the radial wave functions
in (\ref{section5 equation 30}) and (\ref{section5 equation 31}) starting from (\ref{section6 equation 4}) and proceeding as above in
the derivation of the bounds on $\rho_\pm$.

The case $M=-j-\frac{1}{2}$ has to be handled with care because when
$2j+1< \nu <2j+2$ we have noted that $\rho_-$   behaves as $(kR)^{-\lambda}$   as $ k\rightarrow0$ and
hence one might naively conclude that $\psi_{EjMm,-\frac{1}{2}}^-$      behaves as $(kR)^{-\lambda -1}$   with
$ 0< \lambda <1$ when $k\rightarrow 0$. This would induce a non-integrable singularity
in the chiral propagator (\ref{section6 equation 2}). This does not happen for the
following reason. It may be explicitly checked that  $\rho_{0-}$  in (\ref{section6 equation 20})
is given by
\begin{eqnarray}
\rho_{0-} (r) = C r^{2j+\frac{3}{2}}  e^{ - \int_{r_0}^r ds\, sa\, (s) };
\label{section6 equation 21}
\end{eqnarray}
that is, it is a regular solution of (\ref{section5 equation 7}) when $M=-j- \frac{1}{2}$  and $k=0$.
 Here $C$ and $r_0$  are arbitrary constants. This together with (\ref{section6 equation 20})
show that the relevant radial wave function in (\ref{section5 equation 31}) satisfies
\begin{eqnarray}
\frac{1}{k} \left ( \frac{d}{dr} + a r - \frac{2j + \frac{3}{2}}{r}  \right ) \rho_- (r) = O(kR)^{1 - \lambda}.
\label{section6 equation 22}
\end{eqnarray}
Hence, $\psi_{EjMm,-\frac{1}{2}}^-$      continues to vanish as $k\rightarrow 0$ for $M=-j-\frac{1}{2}$   and $0 \leq r \leq R$.
Another potential non-integrable singularity from $\rho_+$  in the second 
term in (\ref{section5 equation 30}) is averted by the vanishing of the Clebsch-Gordon
coefficient at $M=-j- \frac{1}{2}$. In view of the foregoing it is clear that
the detailed analysis here is necessary.

It is now required to examine the low energy behavior of the
radial wave functions in (\ref{section5 equation 30}) and (\ref{section5 equation 31}) when $r >  R$. We choose to
deal with the most troublesome cases in Sec.VI first, namely those
arising when $M = -j-\frac{1}{2}$. We need only consider $\rho_-$   in this case.
For cases 2.1 and 2.3 when  $\lambda = 2j+1-\nu$  get from (\ref{section3 equation 3}) and (\ref{section5 equation 13})
\begin{eqnarray}
\frac{1}{k} \left( \frac{d}{dr} +  ar - \frac{2j + \frac{3}{2} }{r} \right) \rho_- (r)
&=& -\sqrt{\frac{r}{2}} \left(
J_{2j+2-\nu} (kr) \cos \triangle_{j,-j-\frac{1}{2},-\frac{1}{2}}^+ \right. \nonumber \\
&&  - \left . Y_{2j+2 - \nu} (k r) \sin \triangle_{j, -j-\frac{1}{2}, -\frac{1}{2}}^+ \right ).
\label{section6 equation 23}
\end{eqnarray}
For cases 2.2 and 2.4 when $\lambda =\nu  -2j-1$ then
\begin{eqnarray}
\frac{1}{k} \left( \frac{d}{dr} +  ar - \frac{2j + \frac{3}{2}}{r} \right) \rho_- (r)
&=& \sqrt{\frac{r}{2}} \left(
J_{\nu - 2j- 2} (kr) \cos \triangle_{j,-j-\frac{1}{2},-\frac{1}{2}}^+ \right. \nonumber \\
&&- \left . Y_{\nu - 2j-2} (k r) \sin \triangle_{j, -j-\frac{1}{2}, -\frac{1}{2}}^+ \right ).
\label{section6 equation 24}
\end{eqnarray}
Cases 2.5 and 2.7 are obtained by setting  $\nu = 2j+1$ and $2j$, respectively,
in (\ref{section6 equation 23}); case 2.6 is obtained from (\ref{section6 equation 24}) by setting $\nu  = 2j+2$. Using
(\ref{section5 equation 20})-(\ref{section5 equation 24}) and $Y_\rho (z) \sim  -\Gamma  (\rho )(z/2)^{-\rho}  / \pi$  for $z\rightarrow 0$ obtain the
following results with $\alpha$  denoting $E,j,-j-\frac{1}{2} ,m, -\frac{1}{2}$
\begin{eqnarray}
\mbox{case 2.1:  } \psi_\alpha^- &  = &  O(kR)^{\lambda + 1}   , 0< \lambda <1 ,\nonumber \\
\mbox{case 2.2:  } \psi_\alpha^- &  = &  O(kR)^{1 - \lambda}   , 0< \lambda <1, \nonumber \\
\mbox{case 2.3:  } \psi_\alpha^- &  = &  O(kR)^{\lambda+1}   ,   \lambda>1, \nonumber \\
\mbox{case 2.4:  } \psi_\alpha^- &  = &  O(kR)^{\lambda -1}   ,   \lambda >1,              \nonumber \\        
\mbox{case 2.5:  } \psi_\alpha^- &  = &  O(kR), \nonumber \\
\mbox{case 2.6:  } \psi_\alpha^- &  = &  O(1), \nonumber \\
\mbox{case 2.7:  } \psi_\alpha^- &  = &  O(kR)^2.
\label{section6 equation 25}
\end{eqnarray}
For case 1, $M=j+ \frac{1}{2}$, only $\psi_{Ej,j+\frac{1}{2}, m, \frac{1}{2}}^-$ is relevant and
\begin{eqnarray}
\frac{1}{k} \left ( \frac{d}{dr} - ar - \frac{2j + \frac{3}{2}}{r} \right ) \rho_+ & = & - \sqrt{\frac{r}{2} } \left (J_{2j+2 + \nu} (kr) \cos \triangle_{j, j+\frac{1}{2}, \frac{1}{2}}\right. \nonumber \\
&& \left. - Y_{2j + 2 + \nu} (kr) \sin \triangle_{j,j+\frac{1}{2}, \frac{1}{2}}^+ \right ).
\label{section6 equation 26}
\end{eqnarray}
Thus $\psi_{Ej,j+\frac{1}{2}, m, \frac{1}{2}}^-        = O(kR)^{\lambda + 1}$,  $\lambda>1$. Finally, when $|M|\not=j+\frac{1}{2}$   and therefore $\lambda  >1$,
$\psi_{Ej, j+\frac{1}{2}, m, \frac{1}{2}}        = O(kR)^{\lambda - 1}$   based on (\ref{section5 equation 13}), (\ref{section5 equation 18}) and the form of the radial wave functions
appearing in (\ref{section5 equation 30}) and (\ref{section5 equation 31}). All of these cases are for $r \underset{\sim}{>} R$.

Now return to the last term in (\ref{section2 equation 13})  and the interacting propagator
(\ref{section6 equation 2}). As noted earlier the centrifugal barrier term in (\ref{section5 equation 7}) together
with the regularity assumptions made on $a(r)$ will cause the large
$j>> \nu$  contributions to $\triangle_-  (x,x')$ to approach those of the noninteracting
propagator. Therefore we need only consider a finite range of $j$ in
the search for a possible mass singularity in $\triangle_-$ that would result in
a non-vanishing remainder at $m=0$.

For $r>>R$ the radial wave functions in (\ref{section5 equation 30}) and (\ref{section5 equation 31}) are
seen from (\ref{section5 equation 12}) to behave as
\begin{eqnarray}
\frac{1}{k} \left ( \frac{d}{dr} \pm ra + \frac{2j + \frac{1}{2}}{r} \right ) \rho_\pm &=& - \sqrt{ \frac{1}{\pi k} } \left[ 
	\sin  \left(kr - \begin{matrix}\frac{1}{2}\end{matrix} \pi (2j + 1) + \delta_{jM, \pm \frac{1}{2}}^+ - \begin{matrix}\frac{1}{4}\end{matrix}\pi \right) \right. \nonumber \\
&& + \left. O\left ( \frac{\cos (kr)}{kr}  \right ) \right ],
\label{section6 equation 27}
\end{eqnarray}
with $k>>1/r$, and similarly for the other group of wave functions.
Therefore the leading large-distance behavior of these wave functions
is the same as in the noninteracting case except for phase shifts.

We have shown that the low energy wave functions   $\psi_{EjMm,\pm \frac{1}{2}}^-    = O(kR)^{|\lambda -1|} $  ,
or less, for $kR<<1$ in the region between $r=0$ and $r \underset{\sim}{>} R$. Since
$\lambda^2     = (2j+1)^2  + 4M \nu +  \nu^2 $, then for $\nu  >0$ and $M\geq 0$, $ \lambda > 2j+1$ and
$\psi_{EjMm,\pm\frac{1}{2}}        = O(kR)^{2j+\epsilon} $, $\epsilon  >0$, which cannot lead to mass singularities in the last term in (\ref{section2 equation 13})  
that would result in a non-vanishing limit at $m=0$. Hence we restrict
our discussion to the case $M<0$, in particular, the extreme case $M=-j-\frac{1}{2}$.

In (\ref{section6 equation 25}) the largest deviation from the noninteracting case
is case 2.6 when $\nu   = 2j+2$. Focus on this mode in $\triangle_-$. This mode first
opens up when  $\nu  = 2$, the threshold value of $\nu$   for the formation of the
first square-integrable zero mode in the positive chirality sector
according to the discussion following (\ref{section4 equation 9}). From (\ref{section6 equation 2}), the second term
in (\ref{section5 equation 31}), (\ref{section6 equation 24}) and (\ref{section5 equation 25}) one obtains for the worst case $r, r' > R$
\begin{eqnarray}
\triangle_{-}^{j=\frac{1}{2}\nu - 1} (x, x') 
& = & \frac{M (\hat{x}, \hat{x}')}{r r'} \int_0^{\Lambda^2} \frac{dk^2}{k^2 +m^2} 
\left [ 
	J_0 (kr) J_0 (kr') - \frac{\pi}{2} \left( Y_0 (kr) J_0 (kr') 
\right. 
\right. \nonumber \\
&& \left. \left. 
	+ Y_0(kr') J_0(kr) 
\right ) 
\Big /  \ln \left( \frac{kR}{2} \right) + \frac{\pi^2}{4} Y_0(kr) Y_0(kr') \Big / \ln^2 \left( \frac{kR}{2} \right) 
\right] + R_\Lambda, \nonumber \\
\label{section6 equation 28}
\end{eqnarray}
with   $\Lambda R<<1$. The $k-$independent matrix $M$ is obtained from the second
term in (\ref{section5 equation 31}) in the calculation of $\psi_{E\alpha}^{-}   (x) \psi_{E\alpha}^{-\dagger}   (x')$, and $R_\Lambda$  is the
contribution to  $\triangle_-^j$  from the region $k>\Lambda$. The most singular term in $m$
in (\ref{section6 equation 28}) occurs in the first integral, written as
\begin{eqnarray}
\int_0^\infty dk \frac{kJ_0 (kr) J_0 (kr') } { k^2 + m^2} - \int_\Lambda^\infty dk \frac{k J_0 (kr) J_0 (kr') } { k^2 + m^2 } \nonumber \\
= I_0 (mr_<) K_0 (mr_>) - \int_\Lambda^\infty (\cdot) , 
\label{section6 equation 29}
\end{eqnarray}
where $r_< (r_> )$  denotes the lesser (larger) of $r,r'$. The last integral
in (\ref{section6 equation 29}) can be put into the remainder $R_\Lambda$  in (\ref{section6 equation 28}). Then for $m \rightarrow 0$
the most singular behavior in $m$ of  $\triangle_- (x,x')$ occurs in the mode
$j= \frac{1}{2}\nu   -1$ which has only a logarithmic mass singularity when
$r,r' \underset{\sim}{>} R$
\begin{eqnarray}
\triangle_{-}^{j=\frac{1}{2}\nu -1} (x, x') = -2 \frac{M (\hat{x}, \hat{x}')}{r r'} \ln (mr_>) + \mbox{less singular in m}.  
\label{section6 equation 30}
\end{eqnarray}

In summary, a mode-by-mode analysis of the exact propagator
in (\ref{section6 equation 2}) uncovers only minor deviations from the free propagator
in the low energy domain. 
If $\Pi_4^-$ is finite at $m=0$ so that $\underset{m=0}{\lim} m^2 \partial \Pi_4^- / \partial m^2= 0$ and if the role of the symmetry of $F_{\mu \nu}$ at large distances in reaching this conclusion is well-understood then it should be possible to generalize this fourth-order result to $m^2 \partial \Pi_6^- / \partial m^2 $, etc., obtained by expanding $\Delta_-$ in (\ref{section2 equation 13}) in a power series.  
We have shown in this section that its expansion is justified, considering that no nonperturbative singularities are induced in $\Delta_-$ by the scattering states that would cause the $m=0$ limit of the last term in (\ref{section2 equation 13}) to be nonvanishing.  
At this stage we do not see any other feasible way of demonstrating the vanishing of the last term in (\ref{section2 equation 13}) at $m=0$ since the available evidence relies on explicit gauge invariance and the long-distance symmetry of $F_{\mu \nu}$.  


\section{Large-mass Limit of $\mathcal{R}$}
The leading term in the asymptotic expansion of $\mathcal{R}$ in (\ref{section2 equation 10})
for large $m$ can be calculated from the effective Lagrangian
density for QED$_4$  in a constant field background \cite{bib_14, bib_22, bib_21}. 
This
is possible provided $F_{\mu \nu}$   is assumed to be smooth enough so that
a meaningful derivative expansion of $\mathcal{R}$ can be carried out. Just
how smooth will be made more precise below.

The photon-photon scattering graph in $\mathcal{R}$ has been thoroughly
studied by Karplus and Neuman \cite{bib_23}. 
Using their results or the
comprehensive review of the Heisenberg-Euler effective Lagrangian
by Dunne \cite{bib_24} 
one gets for large mass
\begin{eqnarray}
\Pi_4  &=& \frac{1}{5760\pi^2 m^4} \int d^4 x 
\left[ 
4 (F_{\mu \nu} F_{\mu \nu} )^2 - 7 ( {}^{\phantom{1}{*}}F_{\mu \nu} F_{\mu \nu})^2 
\right ] \nonumber \\
&& +O\left ( \frac{1}{m^6} \int d^4 x F_{\alpha \beta} F_{\alpha \beta} F_{\mu \nu} \partial^2 F_{\mu \nu},  
\frac{1}{m^6} \int d^4 x F_{\mu \alpha } \partial_\lambda F_{\alpha \nu} \partial_\lambda F_{\nu \beta}F_{\beta \mu} 
\right ). \nonumber \\
\label{section7 equation 1}
\end{eqnarray}
The sixth-order graph obtained from the expansion of $\ln\det_5$  is of
order $1/m^8$. We now seek the conditions under which the leading term
in $\mathcal{R}$ as $m \rightarrow \infty$, the right-hand side of (\ref{section7 equation 1}), becomes positive. 

\begin{figure}[tbp]
\begin{center}\includegraphics[width=8cm]{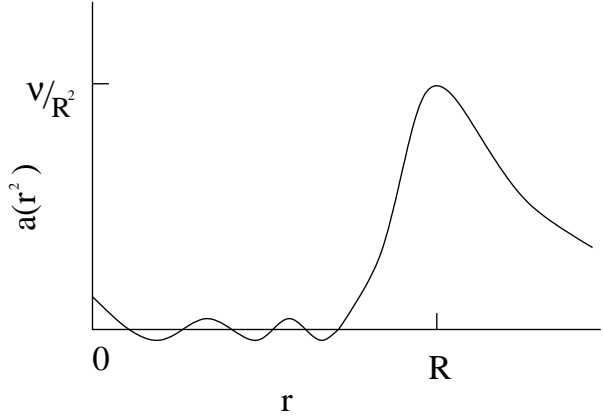}
\caption{
Sketch of $a(r^2)$ versus radial distance for a class of gauge fields satisfying conditions (\ref{section7 equation 4}) and (\ref{section7 equation 5})
}
\label{figure 1}
\end{center}
\end{figure}

From (\ref{section7 equation 1}) positivity requires
\begin{eqnarray}
\int d^4 x (F_{\mu\nu} F_{\mu \nu} )^2 > \frac{7}{4} \int d^4 x ( {}^{\phantom{1}{*}}F_{\mu\nu} F_{\mu \nu})^2.  
\label{section7 equation 2}
\end{eqnarray}
From  (\ref{section3 equation 17}) and (\ref{section3 equation 3}), ${}^{\phantom{1}{*}}F_{\mu\nu} F_{\mu\nu}  = 0$ for $r>R$ and so by (\ref{section3 equation 18}),
$F_{\mu \nu} F_{\mu \nu} = 8\nu^2 / r^4 $ for $r>R$. Then (\ref{section7 equation 2}) gives
\begin{eqnarray}
\int_{r < R} d^4 x (F_{\mu \nu} F_{\mu \nu})^2 + \frac{32 \pi^2 \nu^4}{R^4} > \frac{7}{4} \int_{r < R} d^4 x ( {}^{\phantom{1}{*}}F_{\mu\nu} F_{\mu\nu})^2. 
\label{section7 equation 3}
\end{eqnarray}
Referring again to (\ref{section3 equation 17}) and (\ref{section3 equation 18}) there follows the positivity
condition
\begin{eqnarray}
\int_0^{R^2} dr^2 r^6 a^{'2} [r^4a^{'2} + 4r^2 a a' + a^2 ] > \frac{\nu^4}{R^4},
\label{section7 equation 4}
\end{eqnarray}
where a prime continues to denote differentiation with respect to
$r^2$. It is evident from (\ref{section7 equation 4}) that one class of fields satisfying
the positivity condition is characterized by a steep rise in a in the
region $r \underset{\sim}{<}  R$ where it obtains a maximum before descending
as   $\nu/r^2$  in order to join smoothly with $a(r^2 )$ at $r=R$.
Such an $a$ is illustrated in Fig. \ref{figure 1}. The class of admissible fields
may be larger than this.

In order for the remainder term in the asymptotic expansion
in (\ref{section7 equation 1}) to be finite it is necessary that $F_{\mu\nu}$  be twice differentiable.
From (\ref{section3 equation 1}) the most singular term in $F_{\mu \nu}$ contains terms like $x_\nu M_{\mu \alpha}  x_\alpha a'$
and hence the most singular term in  $\partial^2 F_{\mu\nu}$  is of the form $r^2 x_\nu M_{\mu \alpha}  x_\alpha a'''$.
Thus, the finiteness of  $ \int F^2 F \partial^2  F$ requires
\begin{eqnarray}
\bigg| \int_0^R dr r^7 \left ( \frac{da}{dr} \right )^3 \frac{d^3a}{dr^3} \bigg| < \infty, 
\label{section7 equation 5}
\end{eqnarray}
and so $a(r^2)$ must be at least three times differentiable. For ease
of analysis we assumed in Sec. V that $a(r^2 )$ was regular at the origin,
but this is not necessary. Condition (\ref{section7 equation 5}) only requires
$a\underset{r\rightarrow0}{\sim}    Cr^{\beta}$  with   $\beta > -\frac{1}{2}$. Of course requiring $A_\mu  \in   \underset{n > 4}{\bigcap} L^n (\mathbb{R}^4  )$ rules out
$\beta<0$. Any branches in $a(r^2 )$ away from $r=0$ of the form
\begin{eqnarray}
a (r^2) \underset{r \rightarrow r_0}{\sim} C (r^2 - r_0^2)^\alpha, 
\label{section7 equation 6}
\end{eqnarray}
must have $\alpha  > 5/4$ according to (\ref{section7 equation 5}).

    Now it may happen that a given $a(r^2)$ does not satisfy condition
(\ref{section7 equation 4}). This could mean that either there are no mass
zeros in the remainder defined by (\ref{section2 equation 10}) or that there are an
even number of such zeros. This cannot be decided here. In our
search for definite information we go back to (\ref{section2 equation 10}) and deal only
with $\ln\det_5$, treating the photon-photon graph as a subtraction
like the second-order graph. 
If (\ref{newsection4 equation 2}) is true then the $\ln m^2$ singularity is from 
$\ln\det_5$   alone. Then if the
leading term in $\ln\det_5$'s asymptotic expansion in powers of $1/m$ is
positive it certainly has at least one mass zero in the
interval $0<m<\infty$.

    The leading term in the expansion of $\ln\det_5$  in powers of $1/m$
is the sixth-order graph given by \cite{bib_14, bib_21, bib_22, bib_24}
\begin{eqnarray}
{\ln\det}_5 &=& \frac{1}{40320\pi^2 m^8} \int d^4 x \left [ 
13 ( {}^{\phantom{1}{*}}F_{\mu\nu} F_{\mu\nu} )^2 - 8 (F_{\mu\nu} F_{\mu\nu})^2 
\right ] F_{\alpha \beta} F_{\alpha \beta} \nonumber \\
&&+ O \left(
\frac{1}{m^{10}} \int d^4 x F^2 F^2 F_{\mu \nu} \partial^2 F_{\mu \nu}, \;  \; \frac{1}{m^{10}} \int d^4 x F_{\mu_1 \mu_2} \partial^2 F_{\mu_2 \mu_3} \dots F_{\mu_6 \mu_1} 
\right), 
\label{section7 equation 7}
\end{eqnarray}
and hence the positivity condition is
\begin{eqnarray}
\int d^4 x \left [ 
13 ( {}^{\phantom{1}{*}} F_{\mu \nu} F_{\mu \nu} )^2 - 8 (F_{\mu \nu} F_{\mu \nu})^2
\right ]  F_{\alpha \beta} F_{\alpha \beta} > 0. 
\label{section7 equation 8}
\end{eqnarray}
From (\ref{section3 equation 17}) and (\ref{section3 equation 3}) this becomes
\begin{eqnarray}
\int_{r < R}  d^4x 
\left [
13 ( {}^{\phantom{1}{*}} F_{\mu \nu} F_{\mu \nu} )^2 - 8 (F_{\mu \nu} F_{\mu \nu})^2
\right]F_{\alpha \beta} F_{\alpha \beta} 
> \frac{1024 \pi^2 \nu^6 }{R^8}.
\label{section7 equation 9}
\end{eqnarray}
Further use of (\ref{section3 equation 17}) and (\ref{section3 equation 3}) results in the final positivity
condition
\begin{eqnarray}
\int_0^{R^2} dr^2 \left [
2r^{14} a^{'6} + 12 r^{12}a a^{'5} + 23 r^{10} a^2 a^{'4} + 12 r^8 a^3 a^{'3} 
- 19 r^6 a^4 a^{'2} 
\right ] < \frac{9\nu ^6}{2R^8}.
\label{section7 equation 10}
\end{eqnarray}

    The most singular terms in the remainder of the asymptotic
expansion in (\ref{section7 equation 7}) will arise from those containing  $\partial^2  F$. Following
the above discussion these will be finite provided
\begin{eqnarray}
\bigg | \int_0^R dr r^9 \left ( \frac{da}{dr} \right )^5  \frac{d^3a}{dr^3} \bigg | < \infty,
\label{section7 equation 11}
\end{eqnarray}
This requires $a \underset{r \rightarrow 0}{\sim}    Cr^\beta$    with  $\beta >-1/3$, at least, and any branch points
in a of the form (\ref{section7 equation 6}) must have  $\alpha > 7/6$.

\begin{figure}[tbp]
\begin{center}\includegraphics[width=8cm]{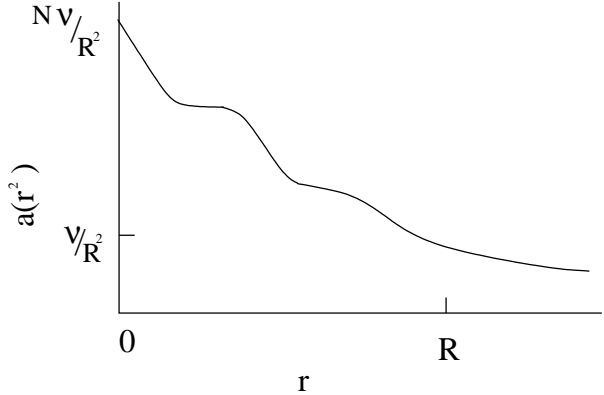}
\caption{
Sketch of $a(r^2)$ versus radial distance for a class of gauge fields satisfying conditions (\ref{section7 equation 10}) and (\ref{section7 equation 11})
}
\label{figure 2}
\end{center}
\end{figure}

    A necessary condition for positivity can be easily derived
from H\"older's inequality, namely
\begin{eqnarray}
\int d^4 x \big |  f g \big | &\leq& \left ( \int d^4 x | f |^p \right )^{\frac{1}{p}} \left ( \int d^4x | g | ^q \right )^{\frac{1}{q}} , \nonumber \\
\frac{1}{p}+ \frac{1}{q} &=& 1, \; \; p, q \geq 1.   
\label{section7 equation 12}
\end{eqnarray}
Then with summation over indices understood, $f=F^2$, $g=( {}^{\phantom{1}{*}}FF)^2$, $p=3$, $q=\begin{matrix}\frac{3}{2}\end{matrix}$,
\begin{eqnarray}
\int d^4 x \left ( 
{}^{\phantom{1}{*}} FF \right)^2 F^2 \leq  
\left( \int d^4 x (F^2)^3 \right )^\frac{1}{3} 
\left( \int d^4 x | {}^{\phantom{1}{*}} FF |^3 \right )^\frac{2}{3}  ,
\label{section7 equation 13}
\end{eqnarray}
and so
\begin{eqnarray}
\int d^4 x \left [ 13 ( {}^{\phantom{1}{*}}FF )^2 F^2 \right. &-& \left. 8 (F^2)^3 \right ] 
\nonumber \\
&\leq&  \left ( \int d^4 x (F^2)^3\right )^\frac{1}{3} 
\left [13 \left ( \int d^4 x | {}^{\phantom{1}{*}} FF |^3 \right )^\frac{2}{3} - 8 \left ( \int d^4 x (F^2)^3 \right )^\frac{2}{3} \right ].
\nonumber \\
\label{section7 equation 14}
\end{eqnarray}
Thus, it is necessary that
\begin{eqnarray}
\int d^4 x \bigg | {}^{\phantom{1}{*}} F_{\mu\nu} F_{\mu \nu} \bigg |^3  > \left ( \frac{8}{13} \right ) ^\frac{3}{2} \int d^4 x \left ( F_{\mu \nu} F_{\mu \nu} \right )^3, 
\label{section7 equation 15}
\end{eqnarray}
for (\ref{section7 equation 8}) to be satisfied. It may be seen by inspection of (\ref{section7 equation 10})
that one class of fields satisfying it are those with
$a(0) \sim   N\nu  /R^2  $, $N \underset{\sim}{>} 2$ and more or less monotonically decaying to $\nu  /R^2$
at $r=R$ as sketched in Fig. \ref{figure 2}. Such fields will not satisfy the
the positivity condition (\ref{section7 equation 4}).

    To summarize, when (\ref{newsection4 equation 2}), (\ref{section7 equation 4}) and (\ref{section7 equation 5}) are satisfied the remainder
$\mathcal{R}$ in (\ref{section2 equation 10}) has at least one zero as $m$ varies over the interval
$0<m< \infty$. When (\ref{newsection4 equation 2}), (\ref{section7 equation 10}) and (\ref{section7 equation 11}) are satisfied, $\ln\det_5$   has such a
zero. In this case the entire function in (\ref{section2 equation 3}) somehow manages
to reduce to unity at the mass zero(s).

\begin{eqnarray}
\nonumber
\end{eqnarray}


\section{Conclusion}
By choosing $O(2)\times O(3)$ symmetric background gauge fields we
were able to make some provisional nonperturbative statements
about the behavior of the Euclidean fermionic determinant $\det_{{ren}}$
of QED$_4$  as a function of the fermionic mass. This determinant has
the form
\begin{eqnarray}
{\ln \det}_{{ren}} = \Pi_2 + \Pi_4 + {\ln \det}_5.
\label{section8 equation 1}
\end{eqnarray}
The second-order term contains a charge renormalization subtraction.
The remaining terms are denoted by $\mathcal{R}$   in (\ref{section2 equation 10}). It was assumed
that for $r>R$ the radial profile function $a(r^2)$ in (\ref{section3 equation 1}) takes the
form  $\nu /r^2$  for $r > R$, together with some mild regularity assumptions
for $a(r^2 )$ for $r<R$. With these assumptions $\det_{{ren}}$     is free of all
cutoffs, including the second-order term if on-shell charge
renormalization is used. Then we showed that if the mass singularity of $\mathcal{R}$ as $m \rightarrow 0$ is fully determined by the chiral anomaly, then  $\mathcal{R}$   has
at least one zero as m varies in the interval $0<m<\infty$   , provided
conditions (166) and (\ref{section7 equation 5}) are satisfied. If not, then provided
(172) and (\ref{section7 equation 11}) are satisfied, $\ln \det_5$   has at least one such zero
at which the entire function in (\ref{section2 equation 3}) becomes unity for any fixed coupling
$e$. Then $\ln \det_{{ren}}$    is dominated by  $\Pi_4$   for  $|e| >> 1$, which is
consistent with $\det_{{ren}}$ being an entire function of order four as
discussed in Sec.II. If there is a mass zero such that $\mathcal{R}$ vanishes
then $\ln \det_{{ren}}    = \Pi_2$    at this zero. If the number of mass zeros in
$\mathcal{R}$   or $\ln \det_5$   is even then they will not show up in the analysis
here.

This raises an interesting possibility.  If $e^2  << 1$ then $m$ does not have to be very large to make a meaningful $1/m$ asymptotic expansion.  So, presumably, there are one or more ``small'' mass zeros in  the weak  coupling domain.  

In plain language the result is this: set $e^2 / \hbar c = 1/137\dots$.  Select a gauge field that satisfies (166), (\ref{section7 equation 5}) or (172), (\ref{section7 equation 11}).  Adjust $m$ until a mass zero appears.  If $m$ is the physical fermion mass then it probably does not  coincide with a mass zero.  But if, for the selected gauge field, $m$ is near a zero then we would  expect the remainder $\mathcal{R}$ or $\ln \det_5$ to be anomalously small compared to the sum of the first few graphs in their expansion.  By continuity there should be a class of gauge fields for which the physical coupling and mass coincide exactly with a mass zero.  

In establishing these results we also demonstrated a
    vanishing theorem when the field strength tensor is not
    (anti-)self-dual, namely, that all of the square-integrable
    zero modes of the Dirac operator are of one chirality. This
    is a generalization of the vanishing theorem of Brown,
    Carlitz and Lee \cite{bib_27}. It would be useful to have a general
    vanishing theorem and to understand the physical principles
    underlying it.


    In Sec. VIII it was assumed that the expansion of $\mathcal{R}$   in powers
of $1/m$ is truly an asymptotic one so that the remainder after the
series is truncated is of the order of the first neglected term.
A proof is needed, but for the present it is an assumption
physicists accept provided the background gauge field is smooth
enough.


Most of this paper deals with the question of whether it is indeed true that the leading mass singularity of $\mathcal{R}$ in (\ref{section2 equation 10}) is determined by the chiral anomaly. We have presented evidence that it is.    
It is true for the case of constant $\mathbf{B}$ and $\mathbf{E}$ \cite{bib_16}, but this is a formal result as the determinant has to be made finite by a volume cutoff. 
 And it is also true for the QCD$_4$ determinant in the presence of an instanton
  background \cite{bib_28}.
 It is evident that the analytic, nonperturbative analysis of four-dimensional fermionic determinants is still at an early stage and may yet yield some surprises.  

\pagebreak
\appendix
\setcounter{equation}{0}
\renewcommand{\theequation}{A\arabic{equation}}

\section*{APPENDIX A}
Here we list some properties of the four-dimensional rotation matrix.  Several of the properties listed can be found in the Appendix of \cite{bib_19} from which our conventions and notations are taken.  

Let $\xi= x_0 + ix_3$, $\eta = x_2 + ix_1$.   Then explicitly 
\begin{eqnarray}
D_{m_1m_2}^l (x) = \left [ (l - m_1)! (l -m_2)! (l + m_1)! (l + m_2)! \right ] ^{\frac{1}{2}} 
\sum_{n_1 \dots n_4} \frac{ \xi^{n_1}\eta^{n_2} (-\bar{\eta})^{n_3} (\bar{\xi})^{n_4} } { n_1 ! n_2 ! n_3 ! n_4! }, 
\label{sectionC equation 1}
\end{eqnarray}
where $n_i = 0, 1, \dots $ and satisfy 
\begin{eqnarray}
n_3 + n_4 & = & l + m_1, \nonumber \\
n_3 + n_1 & = & l + m_2, \nonumber \\
n_1 + n_2 + n_3 + n_4  & = & 2l.
\label{sectionC equation 2}
\end{eqnarray}
The $D_{m_1m_2}^l$ are normalized according to (\ref{section3 equation 11}) where we set $r^2 = x_0^2 + x_1^2 + x_2^2 + x_3^2$.  They satisfy
\begin{eqnarray}
D_{m_1m_2}^{l*} = (-1)^{m_1 + m_2} D_{-m_1, -m_2}^l . 
\label{sectionC equation 3}
\end{eqnarray}
A useful relation is 
\begin{eqnarray}
\sum_{m_2 = -l}^l D_{m_1m_2}^{l*} (\hat{x}) D_{m_1 m_2}^l (\hat{x}) = 1.  
\label{sectionC equation 4}
\end{eqnarray}
The $D_{m_1 m_2}^l $  satisfy an addition theorem
\begin{eqnarray}
\sum_{m_1, m_2 = -l}^l D_{m_1m_2}^{l} (\hat{x}) D_{m_1 m_2}^{l*} (\hat{y}) = C_{2l}^1 (\hat{x} \cdot \hat{y}),   
\label{sectionC equation 5}
\end{eqnarray}
where the Gegenbauer polynomial is 
\begin{eqnarray}
C_{2l}^1 (\cos \varphi) = \frac{ \sin ( 2l + 1) \varphi }{ \sin \varphi }.  
\label{sectionC equation 5}
\end{eqnarray}
They also satisfy a completeness relation
\begin{eqnarray}
\frac{1}{2\pi^2} \sum_{l = 0}^\infty (2 l + 1) \sum_{m_1m_2 = -l }^l D_{m_1m_2}^ l (\hat{x}) D_{m_1m_2}^{l*}(\hat{y}) = \delta (\Omega_{\hat{x}} - \Omega_{\hat{y}}).  
\label{sectionC equation 6}
\end{eqnarray}
The raising and lowering operators of the $O(3)$ subgroups give
\begin{eqnarray}
L_+^{(1)} D_{m_1 m_2}^l & = & - \sqrt{ l (l + 1) - m_1(m_1 + 1) } D_{m_1 + 1, m_2}^l, \nonumber \\
L_-^{(1)} D_{m_1 m_2}^l & = & - \sqrt{ l (l + 1) - m_1(m_1 - 1) } D_{m_1 - 1, m_2}^l, \nonumber \\
L_+^{(2)} D_{m_1 m_2}^l & = &  \sqrt{ l (l + 1) - m_2(m_2 + 1) } D_{m_1, m_2 + 1}^l, \nonumber \\
L_-^{(2)} D_{m_1 m_2}^l & = &  \sqrt{ l (l + 1) - m_2(m_2 - 1) } D_{m_1, m_2 - 1}^l. 
\label{sectionC equation 7}
\end{eqnarray}
Other relations we found useful are
\begin{eqnarray}
\sqrt{j+M+\begin{matrix}\frac{1}{2}\end{matrix}} {\xi} D_{M+\frac{1}{2}, m}^j (x) + \sqrt{j - M + \begin{matrix}\frac{1}{2}\end{matrix} } \bar{\eta} D_{M-\frac{1}{2}, m}^j (x) & = & \sqrt{j - m}r^2 D_{M, m+ \frac{1}{2}}^{j - \frac{1}{2}} (x), \nonumber \\
\sqrt{ j-M+\begin{matrix}\frac{1}{2}\end{matrix} } \bar{\xi} D_{M-\frac{1}{2}, m}^j (x) - \sqrt{ j + M + \begin{matrix}\frac{1}{2}\end{matrix}} \eta D_{M+\frac{1}{2}, m}^j (x) & = & \sqrt{j + m} r^2 D_{M, m-\frac{1}{2}}^{j - \frac{1}{2}} (x),  
\label{sectionC equation 8}
\end{eqnarray}
as well as 
\begin{eqnarray}
\frac{\partial}{\partial \xi} D_{M - \frac{1}{2}, m}^j (x)  & = & ( j - M + \begin{matrix}\frac{1}{2} \end{matrix})^{\frac{1}{2}} (j+m)^{\frac{1}{2}} D_{M, m - \frac{1}{2} }^{j - \frac{1}{2} }  (x), \nonumber \\
\frac{\partial}{\partial \eta } D_{M - \frac{1}{2}, m}^j (x) & = & (j - M + \begin{matrix}\frac{1}{2}\end{matrix})^{\frac{1}{2}} (j-m)^{\frac{1}{2}} D_{M, m + \frac{1}{2} }^{j - \frac{1}{2} }  (x) ,\nonumber \\
\frac{\partial}{\partial \bar{\xi}} D_{M + \frac{1}{2}, m}^j (x) & = & (j + M + \begin{matrix}\frac{1}{2}\end{matrix})^{\frac{1}{2}} (j - m )^{\frac{1}{2}} D_{M, m + \frac{1}{2} }^{j - \frac{1}{2} }  (x), \nonumber \\
\frac{\partial}{\partial \bar{\eta}} D_{M + \frac{1}{2}, m}^j (x) & = &  - (j + M + \begin{matrix}\frac{1}{2}\end{matrix})^{\frac{1}{2}} (j + m )^{\frac{1}{2}} D_{M, m - \frac{1}{2} }^{j - \frac{1}{2} }  (x).  
\label{sectionC equation 9}
\end{eqnarray}

\pagebreak
\setcounter{equation}{0}
\renewcommand{\theequation}{B\arabic{equation}}
\section*{APPENDIX B}
In order to calculate the low-energy phase shifts we follow the procedure of approximating the interior wave function by a small $k^2$ expansion about its zero-energy solution.  Thus, (\ref{section5 equation 7}) has the form
\begin{eqnarray}
\left(\frac{d^2}{dr^2} + f_\pm (r)\right) \rho_{\pm \frac{1}{2}} = -k^2 \rho_{\pm \frac{1}{2}}.  
\label{sectionD equation 1}
\end{eqnarray}
At a zero energy
\begin{eqnarray}
\left(\frac{d^2}{dr^2} + f_\pm \right) \rho_{0, \pm \frac{1}{2}} = 0.
\label{sectionD equation 2}
\end{eqnarray}
From (\ref{sectionD equation 1}) and (\ref{sectionD equation 2}) get, after integrating, 
\begin{eqnarray}
\rho_{0, \pm \frac{1}{2}} (r) \rho_{\pm \frac{1}{2}}^{'} (r) - \rho_{\pm \frac{1}{2}} (r) \rho_{0, \pm \frac{1}{2}}^{'} (r) = -k^2 \int_0^r ds \rho_{\pm \frac{1}{2}} \rho_{0, \pm \frac{1}{2}}, 
\label{sectionD equation 3}
\end{eqnarray}
since $\rho_{\pm \frac{1}{2} } \underset{r \rightarrow 0}{\sim} r^{2j + \frac{3}{2}} $ provided $r^2 a \underset{r \rightarrow 0}{ = } 0 $, which we have assumed.  Hence
\begin{eqnarray}
\left ( 
\frac{ r\partial_r \rho_{\pm \frac{1}{2}} } { \rho_{\pm \frac{1}{2}} } 
\right)_R - \left ( 
\frac{ r\partial_r \rho_{0, \pm \frac{1}{2}} } { \rho_{0, \pm \frac{1}{2}} } 
\right )_R = 
- \frac{ 
k^2 R \int_0^R dr \rho_{\pm \frac{1}{2} }  \rho_{0, \pm \frac{1}{2}} 
} { 
\rho_{\pm \frac{1}{2}}{(R)}  
\rho_{0, \pm \frac{1}{2} }{(R)} 
}. 
\label{sectionD equation 4}
\end{eqnarray}
Since $k^2$ is an analytic perturbation of $\rho_{\pm \frac{1}{2}} $ in (\ref{sectionD equation 1}), make the expansion
\begin{eqnarray}
\rho_{\pm \frac{1}{2}} (r) = \rho_{0, \pm \frac{1}{2}} + \rho_{2, \pm \frac{1}{2}} (r) k^2 + O (k^4), 
\label{sectionD equation 5}
\end{eqnarray}
and substitute this in (\ref{sectionD equation 4}) to get the interior logarithmic derivative
\begin{eqnarray}
\left (
\frac{r \partial_r \rho_{\pm \frac{1}{2} } } { \rho_{\pm \frac{1}{2} } }
\right )_R = \gamma_{\pm \frac{1}{2} } - (kR)^2 \Gamma_{\pm \frac{1}{2}} + O (kR)^4.
\label{sectionD equation 6}
\end{eqnarray}
Here
\begin{eqnarray}
\gamma_{\pm \frac{1}{2}}  &= & \left (
\frac{
r \partial_r \rho_{0, \pm \frac{1}{2} } 
}{
\rho_{0, \pm \frac{1}{2} } }
\right )_R, \nonumber \\
\Gamma_{\pm \frac{1}{2}}  &= & 
\frac{
\int_0^R  dr \rho_{0, \pm \frac{1}{2}}^2
} { 
R \rho_{0, \pm \frac{1}{2} }^{2}  (R )}, 
\label{sectionD equation 7}
\end{eqnarray}
where $\gamma_{\pm \frac{1}{2}} $, $\Gamma_{\pm \frac{1}{2} }$ are denoted by $\gamma_\alpha$, $\Gamma_\alpha$, respectively, in Secs.VI and VII with $\alpha$ denoting $j, M, \pm \frac{1}{2}$.  When $M = -j -\frac{1}{2}$, $\rho_{0, -\frac{1}{2}}$ is given by  
(\ref{section6 equation 21})
and when $M = j + \frac{1}{2}$, 
\begin{eqnarray}
\rho_{0, \frac{1}{2}} = C r^{2j+ 3/2} e^{\int_{r_0}^r ds s a } . 
\label{sectionD equation 8}
\end{eqnarray}

Calculation of the low-energy phase shifts then proceeds by equating the interior derivative (\ref{sectionD equation 6}) with the exterior derivative calculated from the small-$k$ expansion of 
\begin{eqnarray}
\rho_{\pm \frac{1}{2} } (r) = \sqrt{\frac{r}{2}} J_\lambda (kr) \cos \Delta_{\pm \frac{1}{2}}^{+}(k) - \sqrt{\frac{r}{2}} Y_\lambda (kr) \sin \Delta_{\pm \frac{1}{2} }^+ (k), 
\label{sectionD equation 9}
\end{eqnarray}
obtained from 
(\ref{section5 equation 13}) and 
(\ref{section5 equation 16}).
\begin{eqnarray}
\nonumber
\end{eqnarray}

\pagebreak

\end{document}